\documentclass[5p,times]{elsarticle}

\usepackage{amsmath}
\usepackage[singlespacing]{setspace}
\usepackage{xcolor}
\usepackage{caption}
\usepackage{subcaption}
\usepackage{booktabs}
\usepackage{ulem}
\usepackage{hyperref}

\newcommand{\editmade}[1]{#1}
\newcommand{\editmadebl}[1]{#1}
\newcommand{\editstrike}[1]{}

\definecolor{codegreen}{rgb}{0,0.6,0}
\definecolor{codegray}{rgb}{0.5,0.5,0.5}
\definecolor{codepurple}{rgb}{0.58,0,0.82}
\definecolor{backcolour}{rgb}{0.95,0.95,0.92}

\usepackage{listings}
\newcommand*\wraptt[1]{\texttt{\expandafter\ttbreak\detokenize{#1}\relax}}
\newcommand*\ttbreak[1]{\ifx\relax#1\else
  \expandafter\ifx\string_#1\string_\allowbreak\else#1\fi
  \expandafter\ttbreak\fi}

\journal{Parallel Computing}

\begin{document}

\begin{frontmatter}

\title{Enabling GPU Accelerated Computing in the SUNDIALS Time Integration Library}

\author[add1]{Cody J. Balos\corref{corauth}}
\ead{balos1@llnl.gov}

\author[add1]{David J. Gardner}
\author[add1]{Carol S. Woodward}
\author[add2]{Daniel R. Reynolds}

\address[add1]{Lawrence Livermore National Laboratory, 7000 East Avenue, Livermore CA 94550}
\address[add2]{Department of Mathematics, Southern Methodist University, Dallas TX 75275}

\cortext[corauth]{Corresponding author.}

\begin{abstract}
As part of the Exascale Computing Project (ECP), a recent focus of development efforts for the SUite of Nonlinear and DIfferential/ALgebraic equation Solvers (SUNDIALS) has been to enable GPU-accelerated time integration in scientific applications at extreme scales. This effort has resulted in several new GPU-enabled implementations of core SUNDIALS data structures, support for programming paradigms which are aware of the heterogeneous architectures, and the introduction of utilities to provide new points of flexibility. In this paper, we discuss our considerations, both internal and external, when designing these new features and present the features themselves. We also present performance results for several of the features on the Summit supercomputer and early access hardware for the Frontier supercomputer, which demonstrate negligible performance overhead resulting from the additional infrastructure and significant speedups when using both NVIDIA and AMD GPUs.
\end{abstract}


\end{frontmatter}



\section{Introduction}
\label{s:intro}


The SUite of Nonlinear and DIfferential/ALgebraic equation Solvers (SUNDIALS) \cite{hindmarsh2005sundials} is a freely available open source library of robust and scalable time integrators for ordinary differential equations (ODEs) and differential-algebraic equations (DAEs) as well as solvers for systems of nonlinear algebraic equations. The library is comprised of six independent packages: CVODE and IDA provide adaptive order and step implicit linear multistep methods for  ODEs and DAEs, respectively, CVODES and IDAS are extensions of CVODE and IDA with forward and adjoint sensitivity analysis capabilities, ARK\-ODE includes adaptive step additive Runge-Kutta methods as well as multirate methods for ODEs, and KINSOL implements solvers for nonlinear systems. The packages are all built on a set of shared vector, matrix, and solver classes and are designed to be easily incorporated into existing application codes. As such, SUNDIALS has been integrated into numerous mathematical libraries and application codes including projects for block-structured adaptive mesh refinement \cite{zhang2020amrex}, high-order finite elements \cite{anderson2021mfem}, cosmology \cite{almgren2013nyx}, combustion \cite{day1999pele, nonaka2017pele}, and materials science \cite{dorr2010numerical}, all targeting exascale computing systems.

The computational power provided by exascale systems will enable these and other application codes to simulate far larger and more detailed models than ever before, incorporating more physical processes and scales, giving rise to increased predictive capabilities. A robust mathematical library infrastructure providing novel algorithms and highly performant implementations is a key component in ensuring the numerical stability, accuracy, and computational efficiency of these complex models. Thus, as applications prepare for the challenges that heterogeneous computing architectures present, i.e., extreme levels of concurrency, deep memory hierarchies, new conceptual and programming paradigms, etc., mathematical libraries must adapt to these challenges as well \cite{karlin2019preparation}.

As SUNDIALS resides in the middle of the typical numerical software stack, both being called by applications or discretization frameworks and calling other mathematical libraries, we must address these challenges in a manner that will accommodate the approaches utilized by application codes and other packages to achieve performance and portability.
Meeting these challenges includes supporting various memory management strategies (distinct host and device memory spaces, unified virtual memory, memory pools, etc.) and programming models (vendor specific languages, open standards, and performance portability layers). Additionally, the SUNDIALS time integration packages need to be able to support vastly different application use cases: (1) evolving large coupled systems of ODEs, and (2) evolving numerous small independent systems.  These use cases each include distinct requirements for algebraic solvers and kernel optimizations critical to performance. Naturally, fulfilling such solver requirements necessitates seamless and performant interoperability with other libraries on heterogeneous systems. Moreover, addressing these challenges must also be done while managing the complexity and ensuring the sustainability of the library.

As detailed in \cite{gardner2020enabling}, we have recently made several enhancements to SUNDIALS as part of the United States Department of Energy's Exascale Computing Project (ECP) to increase the \editmade{modularity and} extensibility of SUNDIALS \editmade{by creating new abstract classes for linear and nonlinear solvers and enhancing the capabilities of the abstract vector class. These class interfaces are briefly summarized below and provide the foundation for new class implementations, described in this work, that address the aforementioned challenges in supporting applications running on hybrid CPU+GPU systems.}
This paper details the design considerations that went into creating the SUNDIALS GPU features, presents the current GPU capabilities, and demonstrates the GPU performance of the SUNDIALS time integrators on a demonstration problem.

In the following section we present an overview of the \editmade{object-oriented} structure of SUNDIALS and the typical GPU use cases that inform the design of GPU support in SUNDIALS. Advanced memory management features for GPUs are discussed in Section \ref{s:sunmemory} and are  followed by a presentation of the suite's GPU enabled vector, matrix, and solver structures in Sections \ref{s:vectors} and \ref{s:matrix-solver}. Section \ref{s:usingsundials} highlights key considerations when using SUNDIALS on hybrid CPU+GPU systems. Results from a demonstration code illustrating the SUNDIALS GPU capabilities are given in Section \ref{s:demonstration}. Finally, in Section \ref{s:conclusions} we summarize the state of the SUNDIALS GPU capabilities and identify next steps for development.

\section{Design Considerations}
\label{s:considerations}


As mentioned above, SUNDIALS packages are designed to be utilized by application codes or discretization frameworks and, in turn, SUNDIALS often calls other libraries that provide the algebraic solvers needed for the solution of nonlinear and linear systems that arise within implicit integration methods.
As a result, SUNDIALS occupies a position in the middle of the typical numerical software stack and must accommodate application and framework data layouts and parallelization approaches while simultaneously providing the flexibility for data to be seamlessly transferred to external solvers.
This situation is further complicated in the GPU context by the variety of memory spaces and programming paradigms used by different accelerated hardware. Moreover, the transfer of data and control must be done in a way that allows the solver packages to fully exploit the power of GPUs.

To address these challenges, SUNDIALS utilizes an \editmade{object-oriented} design where \editmade{mathematical operations} are written fully in terms of abstract operations on generic objects or calls to \editmade{user-supplied callback functions}.
As a result, \editmade{most of the core time-integration and solver} algorithms are agnostic to \editmade{underlying data layouts and parallelism}.
This encapsulation relies on four base SUNDIALS classes that define generic vector, matrix, linear solver, and nonlinear solver operations.
In the case of vectors, these operations are either streaming operations like adding two vectors or scaling a vector, or reduction operations, like norms or dot products. Similarly, the generic matrix class defines operations like scaling and adding matrices or computing a matrix-vector product.
The linear and nonlinear solver objects, working in conjunction with the vector and, if necessary, matrix objects, provide generic operations to set up and solve linear and nonlinear systems, respectively.
\editmade{Implementations of these classes may utilizes different algorithms in order to  better leverage the unique features of the target hardware or programming model.
Readers interested in more details on the object-oriented design of SUNDIALS should see \cite{gardner2020enabling}, or the CVODE user documentation \cite{cvodeDocumentation} for information on how users can define their own implementations of the base classes.}

As the specific details for carrying out an operation are completely contained within a class implementation, the SUNDIALS integrators can be utilized within many different parallel contexts simply by providing class implementations tailored to the desired data structure(s) and target architecture(s).
This flexibility provides a means to accommodate many programming models, including those necessary to utilize NVIDIA, AMD, and Intel GPUs with the same integrator source code. As this design approach suggests, the integrator control logic resides on the host (CPU) while the class implementations operate on data that resides in whatever memory space the object dictates. Thus, users are able to deploy approaches best suited to their application use case in terms of memory management and programming paradigm by providing their own class implementations. Alternatively, SUNDIALS provides a variety of native GPU class implementations (see Sections \ref{s:vectors} and \ref{s:matrix-solver}) that are designed to meet the needs of the most common SUNDIALS use cases.

\begin{figure}
    \centering
    \includegraphics[width=0.35\textwidth]{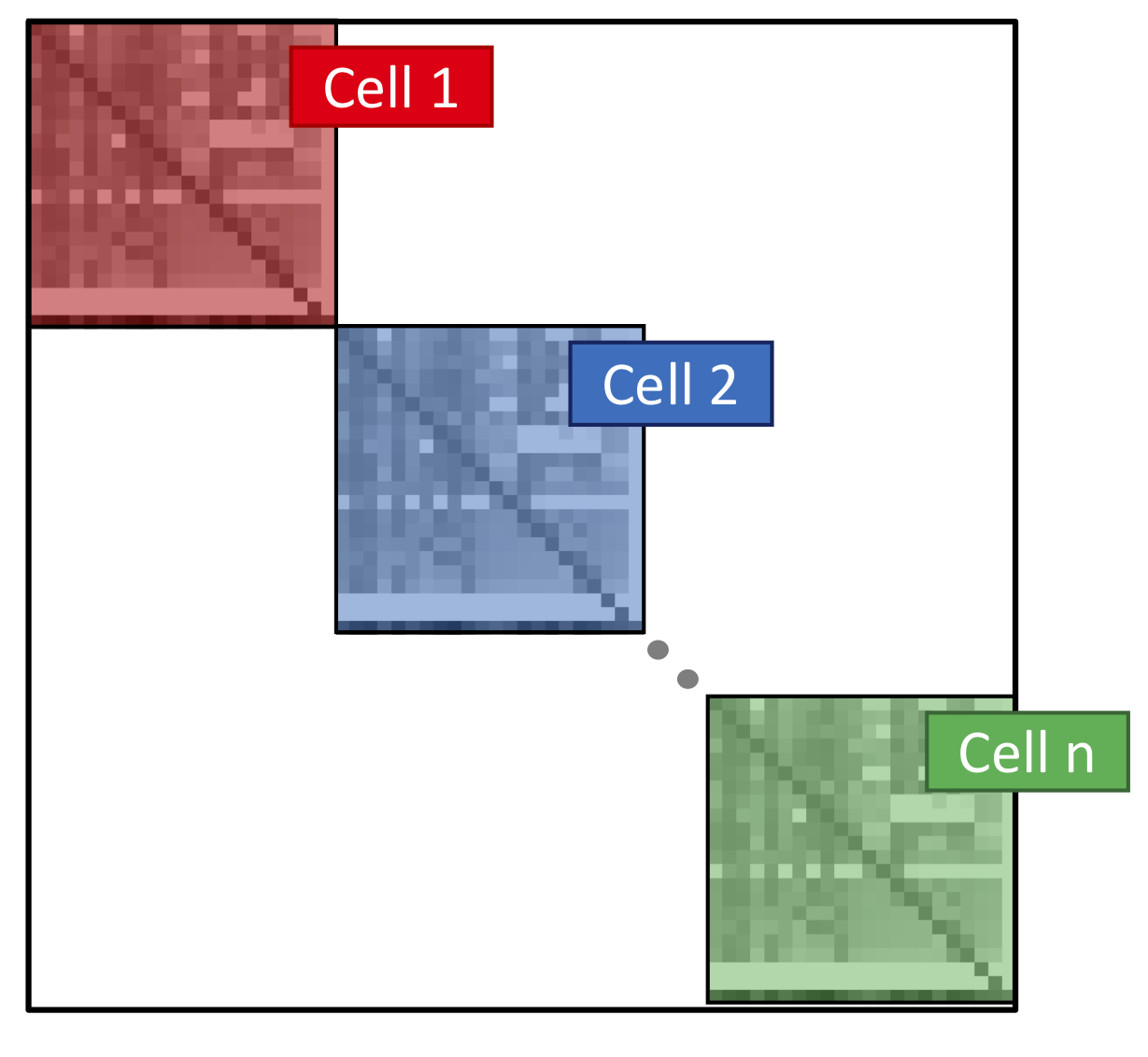}
    \caption{The block diagonal Jacobian formed when grouping many small independent ODEs, occurring in each grid cell of the domain, into a larger ODE.}
    \label{fig:blockdiag_jacobian}
\end{figure}

The SUNDIALS integration packages are primarily used in two contexts. In the first case, a SUNDIALS integrator is invoked to evolve the full application model in time, i.e., advancing the nodes in a large-scale spatially discretized partial differential equation from one time step to the next.
In the second case, a SUNDIALS integrator is used to evolve many small, independent systems within a larger application. This use case, for example, arises in combustion simulations where local chemical kinetics systems must be advanced in each grid cell of the spatial domain.  \editmade{We will refer to these use cases as ``full model'' and ``submodel'' throughout the remainder of this paper, respectively.}

\editmade{In both scenarios}, minimizing data movement between the host and device is critical to performance. As such, object data should remain resident in device memory as much as possible, ideally for the full duration of the simulation. User-defined class implementations fully control their memory management; however, for the native SUNDIALS data structures a minimal interface is provided to allow for greater flexibility in the  allocation, deallocation, and movement of data (see Section \ref{s:sunmemory}).

\editmade{Additionally}, performance hinges on providing a sufficient amount of work to occupy the device.
In the \editmade{full model use} case, saturating the device is less of a concern than in the \editmade{submodel use} case as each independent system on its own may not constitute a sufficient amount of work to efficiently utilize the GPU. In such cases, the small systems can be grouped together into a larger single system and integrated together. \editmade{A caveat is that large variations in stiffness among the small systems can result in overly-restrictive step sizes for the larger system. As such, the performance benefits of this approach can depend on the variation in difficulty between the smaller systems and how the smaller systems are grouped to from the larger system and how well this potential expense is balanced by increased performance with longer vectors.}
When using this approach with implicit integration methods, the resulting Jacobian of the system has a block diagonal structure (for an example see Figure \ref{fig:blockdiag_jacobian}) which can be exploited for greater concurrency in linear solves.
With implicit integration methods and the \editmade{full model} use case, performant general purpose and application-specific algebraic solvers on GPUs are also essential for efficiency.
SUNDIALS includes native solver objects targeting these distinct cases (see Section \ref{s:matrix-solver}) and, for other use cases, SUNDIALS\editmade{' object-oriented} design allows users to provide application-specific or third party solvers as needed.

\editmade{Finally}, we note that SUNDIALS data structures must be able to adjust their device utilization in order to optimize operations for both the large and small local problem sizes that arise in \editmade{either} use case.
To this end, where possible, the native data structures provide a means for setting execution policies for different classes of operations (see Section \ref{s:vectors} and \ref{s:matrix-solver}).

\section{Memory Management}
\label{s:sunmemory}


The memory hierarchies of exascale supercomputers introduce memory management challenges that vary depending on the machine and programming model. In some cases, applications and libraries will explicitly manage data coherency between distinct CPU (host) and GPU (device) memory spaces. Alternatively, they may utilize unified virtual memory (UVM) that is addressable from the host or device. This is often referred to as managed memory, since the programming model and runtimes handle the necessary data migration implicitly. In addition, some programming models allow for other types of memory to be allocated, e.g. CUDA page-locked (pinned) memory, that may be useful in some contexts.

To handle these memory management complexities, many applications provide memory management data structures, or employ a memory management library such as Umpire \cite{beckingsale2020umpire}. To support these applications, SUNDIALS provides a set of utilities collectively referred to as the SUNMemoryHelper API. This API is not designed to be a full featured resource management abstraction like Umpire. Rather the intent is to provide a minimal common interface that the native SUNDIALS data structures (i.e., vector, matrix, and solver objects) can leverage to maximize code reuse and allow for user-provided memory management. \editmade{As such, most} of the GPU-enabled, native SUNDIALS data structures utilize the SUNMemoryHelper API to abstract away the specific details of allocating and deallocating memory, as well as copying data from one memory space to another.

The SUNMemoryHelper API is comprised of the SUNMemory structure and the SUNMemoryHelper class.
The SUNMemory structure encapsulates three fields which are crucial to allowing the SUNMemoryHelper class to define generic operations for allocating, deallocating, and copying memory. These fields are: a void data pointer for the memory wrapped by the structure, a flag indicating ownership of the pointer to keep track of who is responsible for freeing the allocated data (SUNDIALS or the user), and a memory type identifier (host, device, UVM, or pinned).
The ownership flag is needed to allow for user-provided data pointers, and the memory type identifier is used by the SUNMemoryHelper to determine how to copy data from the data pointer in one SUNMemory object to the data pointer in another SUNMemory object which may store data in a different memory space (host vs device).
The SUNDIALS data structures that use the SUNMemoryHelper API hold SUNMemory objects, instead of raw pointers for data arrays, and also hold a SUNMemoryHelper object for performing the necessary memory operations. Natively, SUNDIALS provides \editmade{SUNMemoryHelper implementations that interface to each device-specific language that we support}.

\section{Vectors}
\label{s:vectors}


\begin{figure*}[!htb]
    \centering \includegraphics[width=0.75\textwidth]{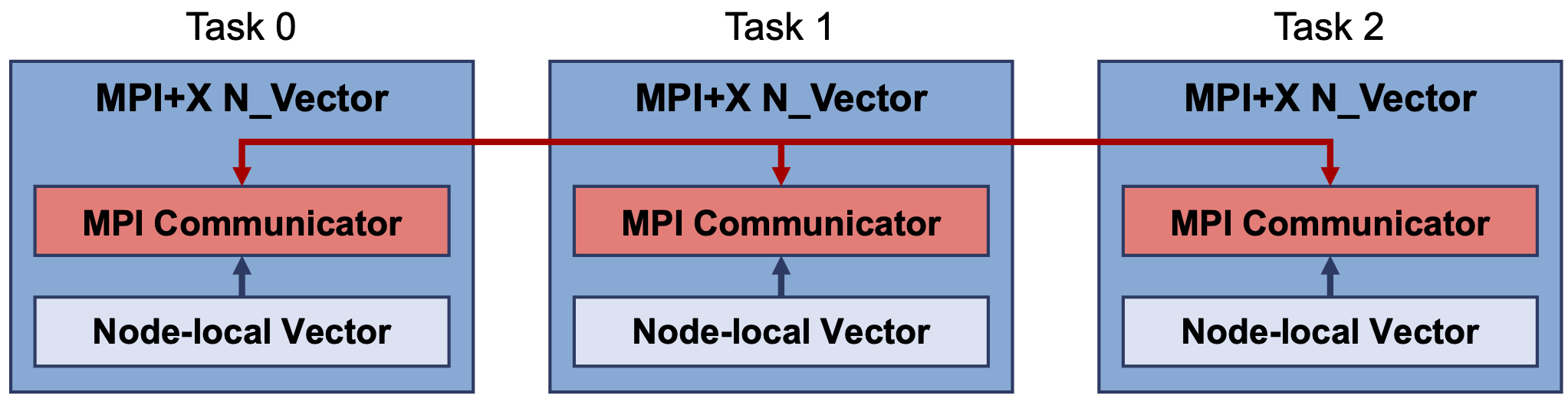}
    \caption{An illustration of the MPI+X vector with three MPI tasks. The MPI+X vector
    contains the MPI communicator and a reference to the node-local vector. The MPI+X
    vector invokes the node-local vector operations and then performs any necessary
    communication between the node-local vectors on each processor.}
    \label{fig:mpi+x}
\end{figure*}

An important architectural feature of modern extreme-scale supercomputers is high compute-density.
Thus, achieving high utilization of the available computational power requires a hybrid parallelism approach.
A common form of hybrid parallelism is the so-called ``MPI+X'' paradigm where cross-node parallelism is achieved with MPI and further on-node parallelism is achieved with some programming model X.
The MPI\-+X approach can be expanded upon to create complex combinations of MPI and multiple forms of on-node parallelism as well.

\editmade{To support such complex combinations}, SUNDIALS provides a ``many-vector'' implementation of the SUNDIALS N\_Vector class, an abstraction layer of a mathematical vector that the time integrators utilize to perform operations on data.
The ManyVector implementation allows for an arbitrarily complex partitioning of vector data across different computational resources and memory spaces, by wrapping $n$ distinct vectors into a single cohesive vector for the integrators and algebraic solvers to operate on.
For further details about the \editmade{abstract N\_Vector base class and the SUNDIALS ManyVector implementation}, readers should refer to \cite{cvodeDocumentation}.

For the simpler MPI+X partitioning, SUNDIALS provides a thin wrapper around the ManyVector, known as the MPIPlusX vector.
Users of SUNDIALS who wish to employ the MPI+X approach to hybrid parallelism create an MPIPlusX vector from any node-local vector object and an MPI communicator to connect the local vectors, as illustrated in Figure \ref{fig:mpi+x}.
The MPIPlusX vector is a specific type of ManyVector where the main vector handles MPI communication between vectors that perform node-local operations.

\editmade{All streaming operations with an MPIPluxX vector use} the on-node parallelism defined by the node-local vector.
In the case of this work, these vectors are enabled with CUDA \editmade{\cite{cuda}}, RAJA \editmade{\cite{beckingsale2019raja}}, \editmade{HIP \cite{hip42manual}}, \editmade{and} OpenMP \editmade{\cite{openmp51manual}} (CPU-only and also with target offloading).
Similarly, for reduction operations the local contributions to the global value are computed using the node-local vector, and the MPIPlusX vector performs the global reduction necessary to obtain the final result.

\editmade{The} MPIPlusX vector is important for the long term sustainability of SUNDIALS, as it removes the need to create nearly identical MPI-parallel and node-local versions of vectors.
For users, the MPIPlusX API, detailed in the SUNDIALS documentation \cite[Section 7.15]{cvodeDocumentation}, provides enough flexibility that switching between different node-local vectors in an application can be done by simply changing which vector constructor function is called, \editmade{e.g. see the code snippet in Listing \ref{lst:mpiplusx_code}} from the
demonstration program discussed in Section \ref{s:demonstration} (simplified to show switching between just two vectors for brevity).
\begin{lstlisting}[linewidth=0.95\columnwidth, breaklines=true, caption={\label{lst:mpiplusx_code} \editmade{The MPIPlusX vector can be used to combine MPI parallelism with on-node parallelism. Any on-node vector can be supplied to the MPIPlusX constructor, and the resulting vector uses both forms of parallelism.}}]
// Local vector length.
int N;

// Create a node-local vector.
N_Vector LocalNvector;
#if defined(USE_GPU)
//   CUDA version.
LocalNvector = N_VNew_Cuda(N);
#else
//   Serial version.
LocalNvector = N_VNew_Serial(N);
#endif

// Create the solution vector.
N_Vector y;
y = N_VMake_MPIPlusX(..., LocalNvector);

// ...
\end{lstlisting}

Several node-local N\_Vector implementations are provided with SUNDIALS. The serial vector, as the name implies, performs all operations on a single CPU core. The Pthreads and OpenMP vectors provide CPU multi-threading, while the OpenMPDEV, CUDA, HIP, and RAJA vectors utilize GPU acceleration to perform the vector operations.
Specifically the OpenMPDEV vector utilizes OpenMP target offloading, the CUDA vector utilizes the NVIDIA CUDA programming model, the HIP vector utilizes the AMD HIP programming model, and the RAJA vector utilizes the RAJA performance portability layer to target CUDA \editmade{or HIP}.
Future work will introduce additional vectors that leverage the \editmade{Kokkos performance portability library \cite{CarterEdwards20143202}} as well as Intel's oneAPI.
While maintaining a plethora of GPU vector implementations is not sustainable for the long-term, SUNDIALS provides these distinct GPU vectors for maximum flexibility during the current era of rapid growth in GPU programming models and architectures.  In the future, once specific GPU programming models have become more established, these node-local implementations may be coalesced.

\subsection{GPU Accelerated Vectors}

A unique feature of the GPU accelerated vectors is that they must handle
different memory management strategies. The CUDA\editmade{, HIP,} and RAJA vectors are designed to utilize distinct host and device memory spaces or unified virtual memory.
This is achieved through the SUNMemory class, discussed in Section \ref{s:sunmemory}, that allocates, deallocates, and moves the vector data appropriately. Therefore, the CUDA\editmade{, HIP,} and RAJA vectors can also leverage \editmade{application-defined memory manager or memory pools.}
The OpenMPDEV vector is currently limited to using distinct host and device memory spaces and employs the OpenMP standard-defined \wraptt{omp_target_alloc} function to allocate device memory and \wraptt{malloc} to allocate host memory.
\editmade{All of the SUNDIALS-provided GPU vectors perform operations on the GPU and require that the data is coherent and accessible on the GPU as a one-dimensional array.}
\editstrike{As such, when using a SUNDIALS-provided GPU vector with distinct host and device memory the user is in complete control of the data coherency, and the SUNDIALS time integrators and solvers will only operate on the data in the device memory space.
The consequences of this are further described in Section \ref{s:usingsundials}}. For access to the underlying data pointers, the GPU vectors define functions of the form
\wraptt{N_VGet<Host|Device>ArrayPointer_*} where the \texttt{*} is the name of the vector implementation.
To move the internal data from one memory space to
another, the vectors define the functions \wraptt{N_VCopy<From|To>Device_*}.

As explained in Section \ref{s:considerations}, the data operations defined by the SUNDIALS N\_Vector class can be grouped into streaming and reduction operations.
For reduction operations, the SUNDIALS GPU vectors perform reductions entirely on the GPU and the scalar result is returned to the host through a memory transfer, thus synchronizing the host and the device.
When possible and applicable, the reduction result is stored in
page-locked (pinned) memory to maximize the efficiency of the transfer.
In contrast, the streaming vector operations are executed asynchronously from the host.
Thus, a transfer of control from SUNDIALS to the user program must be followed by a synchronization of the host and the device before accessing data on the host or outside of the stream of execution used with the GPU vector. \editmade{The listing \ref{lst:synchronize} demonstrates this concept}.
\begin{lstlisting}[linewidth=0.96\columnwidth, breaklines=true, caption={\label{lst:synchronize} \editmade{Streaming vector operations are executed asynchronously with the host thus, a user must ensure the device and host are synchronized before accessing the data on the host. The CUDA code in this listing demonstrates this with an explicit call to a CUDA synchronization function.}}]
// Create a CUDA N_Vector of length 10.
N_Vector y1 = N_VNew_Cuda(10);
N_Vector y2 = N_VNewManaged_Cuda(10);

// Set all entries of the vector y1 to zero.
N_VConst(0.0, y1);

// Copying from device has an implicit synchronize.
N_VCopyFromDevice_Cuda(y1);

// Now we can access the y1 data on the host.
double* data1 =
    N_VGetHostArrayPointer_Cuda(y1);
my_function_that_prints_vector(data1);

// Set all entries of the vector y2 to zero.
N_VConst(0.0, y2);

// Since y2 uses managed memory, we don't need to copy.
// Instead, we explicitly synchronize to ensure N_VConst completed.
cudaDeviceSynchronize();

// Now we can access the y2 data on the host.
double* data2 =
    N_VGetHostArrayPointer_Cuda(y2);
my_function_that_prints_vector(data2);
\end{lstlisting}

\editmade{The CUDA and HIP vector implementations provide} two more notable features, \editmade{not currently supported by the RAJA and OpenMPDEV implementations,} both of which allow users to have further control over how the operations execute on the GPU.
One of these features is the ability to use \editmade{CUDA/HIP} streams. Once the desired stream is set, all of \editmade{the vector} operations, including data movement, will execute in that stream. Thus, \editmade{the vector} operations may be overlapped with other application operations, allowing for greater concurrency as discussed further in Section \ref{s:usingsundials}.
The second feature enables varying the kernel parameters, that is the \editmade{CUDA/HIP} kernel grid and block sizes, to optimize execution for differing application contexts and vector lengths. Therefore, for maximum flexibility, the CUDA \editmade{and HIP vectors also provide} a function \wraptt{N_VSetKernelExecPolicy_*} which takes an an object derived from the \editmade{\wraptt{*ExecPolicy}} class.
This class defines virtual functions that return the kernel parameter values for the grid size, block size, and stream. SUNDIALS natively provides two execution policy implementations for streaming vector operations.
The \editmade{\wraptt{*ThreadDirectExecPolicy}} maps a single GPU thread to a single vector element, while the \editmade{\wraptt{*GridStrideExecPolicy}} maps multiple elements to a GPU thread for a ``grid-stride loop'' strategy.
For reduction operations SUNDIALS provides the \editmade{\wraptt{*BlockReduceExecPolicy}} implementation for block reductions.
Additionally, users may define their own implementation of the \editmade{\wraptt{*ExecPolicy}} to employ alternative strategies for their applications. \editstrike{The HIP vector follows the CUDA vector implementation and provides the same features.}

\subsection{Vector Performance}\label{ss:vectors-perf}

In this section, we present performance results of the native SUNDIALS vectors in isolation by measuring performance of each individual N\_Vector operation on random double precision data.
All tests \editmade{utilize SUNDIALS version 5.6.0 and are} run on the Summit supercomputer at Oak Ridge National Laboratory\editmade{\footnote{All tests were compiled with the IBM XL compiler version 16.1.1-8. The CUDA version used was 10.1.243.   The RAJA version used was v0.12.1}}.
Tests of the node-local vectors are run without MPI, and the GPU vectors utilize one of the NVIDIA V100 GPUs on a Summit compute node.
The OpenMP vector utilizes 42 threads, one per CPU core available on a Summit compute node.

\begin{figure*}[!htb]
    \centering
    \begin{subfigure}[b]{0.49\textwidth}
        \includegraphics[width=\textwidth]{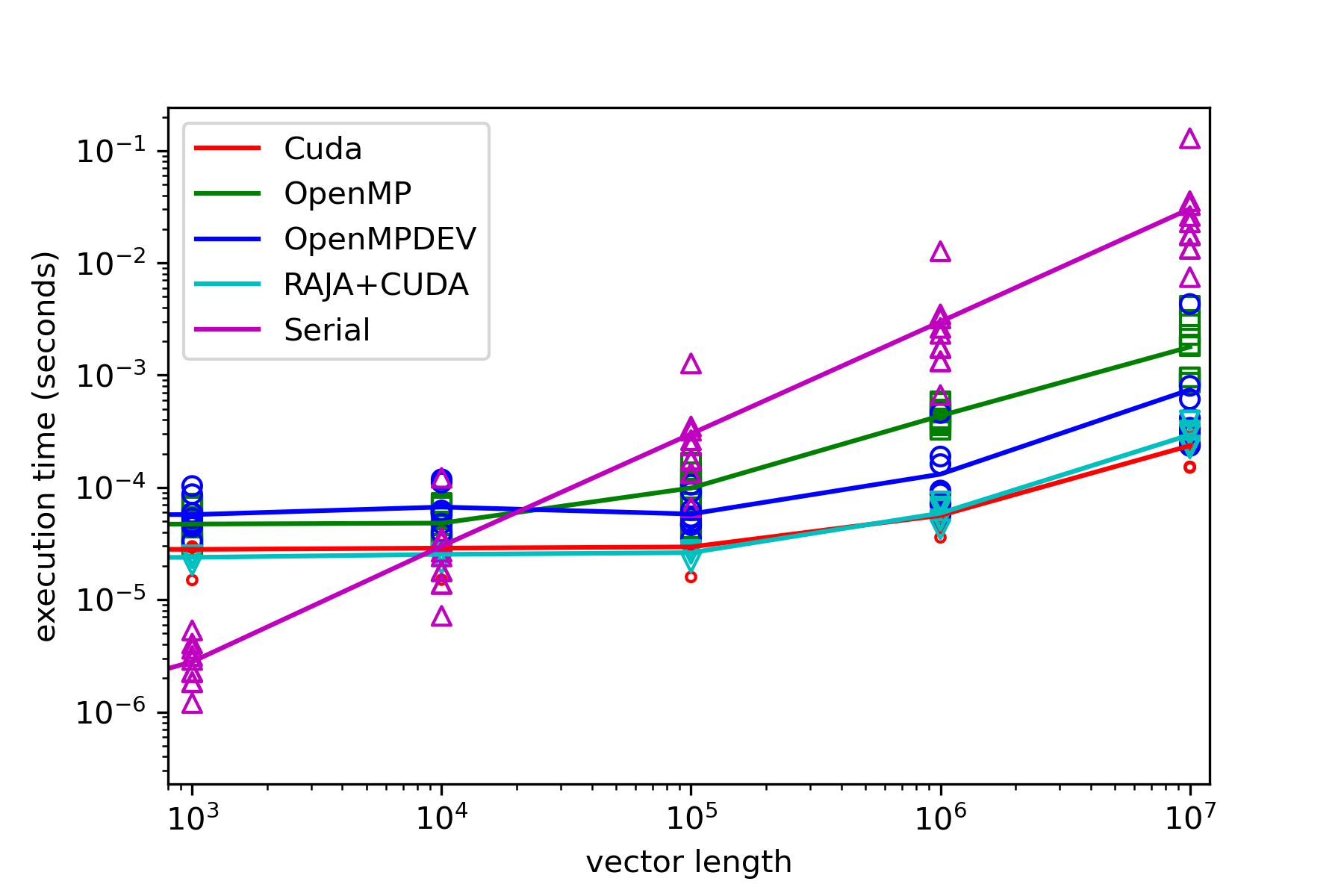}
        \caption{Reduction vector operations}
    \end{subfigure} \hfill \begin{subfigure}[b]{0.49\textwidth}
        \includegraphics[width=\textwidth]{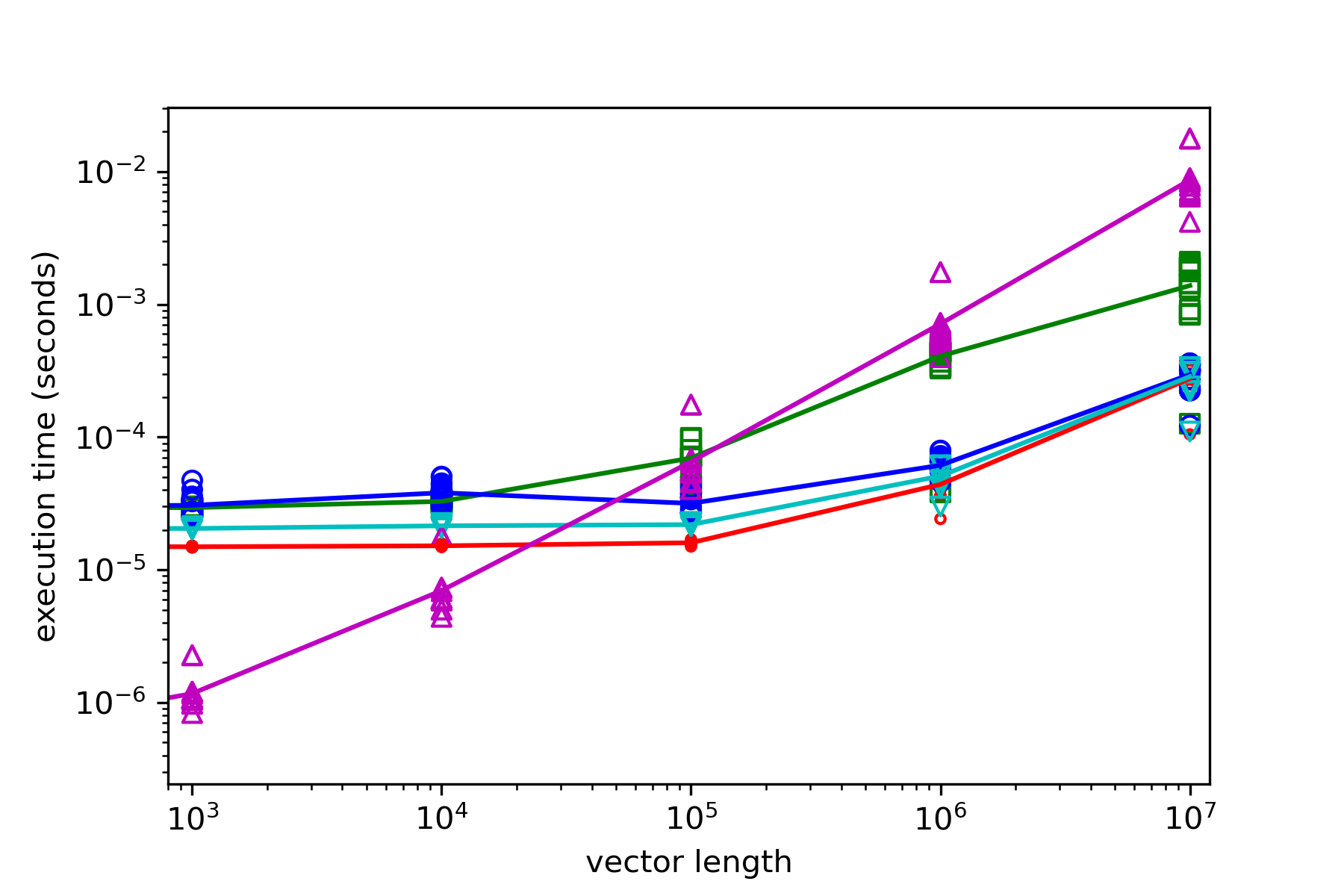}
        \caption{Streaming vector operations}
    \end{subfigure}
    \caption{Node-local performance of various vectors supplied with SUNDIALS.
    Here, a marker represents the mean execution time of a specific vector
    operation using a specific vector implementation on random data. A best-fit
    curve is plotted through each set of markers. We see that the accelerated
    vectors are slower than the serial vector until the vector length exceeds
    $\approx 10^4$.  The maximum speedup (385x) is achieved by the
    \wraptt{N_VConstrMask} reduction operation with the CUDA vector when the
    length is $10^7$. Note, this study does not account for several factors
    that arise in the simulation of a complete problem. See Section \ref{s:demonstration} for performance comparisons in the context of evolving a system of ODEs.}
    \label{fig:onnodeperf}
\end{figure*}

The node-local vector results are plotted in Figure \ref{fig:onnodeperf}, with
all of the streaming vector operations and all of the reduction operations
plotted separately. These results are generated by performing each individual
vector operation on random data 50 times and taking the mean execution time. We
observe that the serial vector is faster than the accelerated vectors until the
vector length reaches $\approx 10^4$. This crossover point is easily explainable: until the
vector reaches a sufficient length, the serial operation execution time is less
than the approximately 8 microseconds it takes to launch the kernel (measured by
launching an empty kernel one hundred thousand times). Furthermore, when using
RAJA and OpenMPDEV, additional kernel launch overhead is observed. Thus, it is unsurprising that the GPU vectors do not show benefit until the serial operation execution time
exceeds $\approx 10^{-5}$ seconds.

\editmade{For} longer vectors, the speedup provided by the
accelerated vectors is significant for both streaming and reduction operations. The
\wraptt{N_VConstrMask} operation achieves the maximum observed speedup of 385x with a
vector that has $10^7$ entries. Since this study performs the vector operations
in isolation, e.g. not as part of the simulation of an ODE or DAE problem, it is important to
note that this study should not be interpreted as the only, or most important,
guidance on which node-local vector to use. In a complete simulation other
factors, such as the required data movement between memory spaces, the expense of
the ODE right hand side function or DAE residual supplied by an application, and the linear solvers
needed by the application play a larger role in determining the optimal vector
choice (as shown in Section \ref{s:demonstration}).

\begin{figure*}[!htb]
    \centering
    \begin{subfigure}[b]{0.54\textwidth}
        \includegraphics[width=\textwidth]{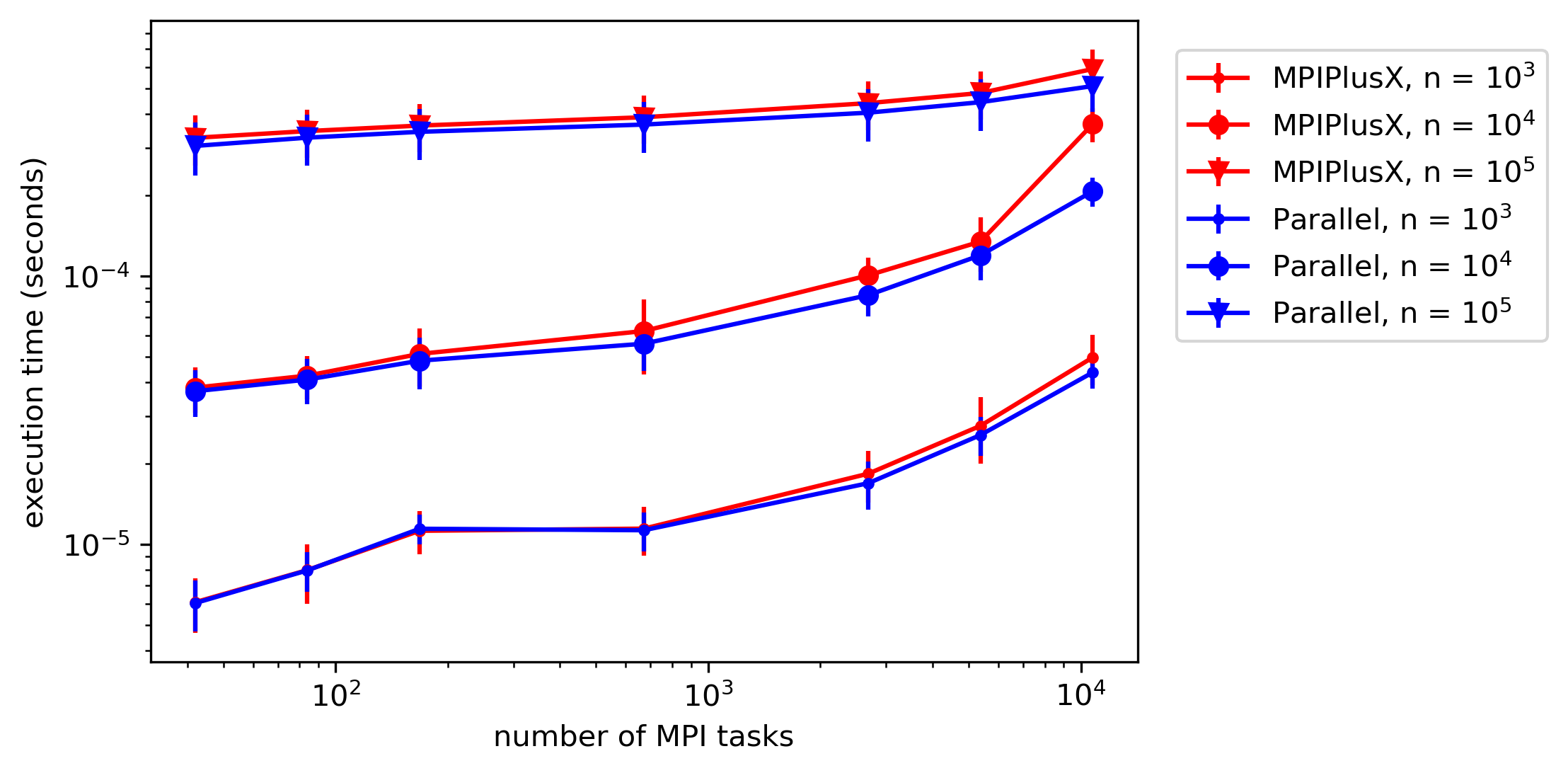}
        \caption{Reduction vector operations}
    \end{subfigure} \hfill \begin{subfigure}[b]{0.44\textwidth}
        \includegraphics[width=\textwidth]{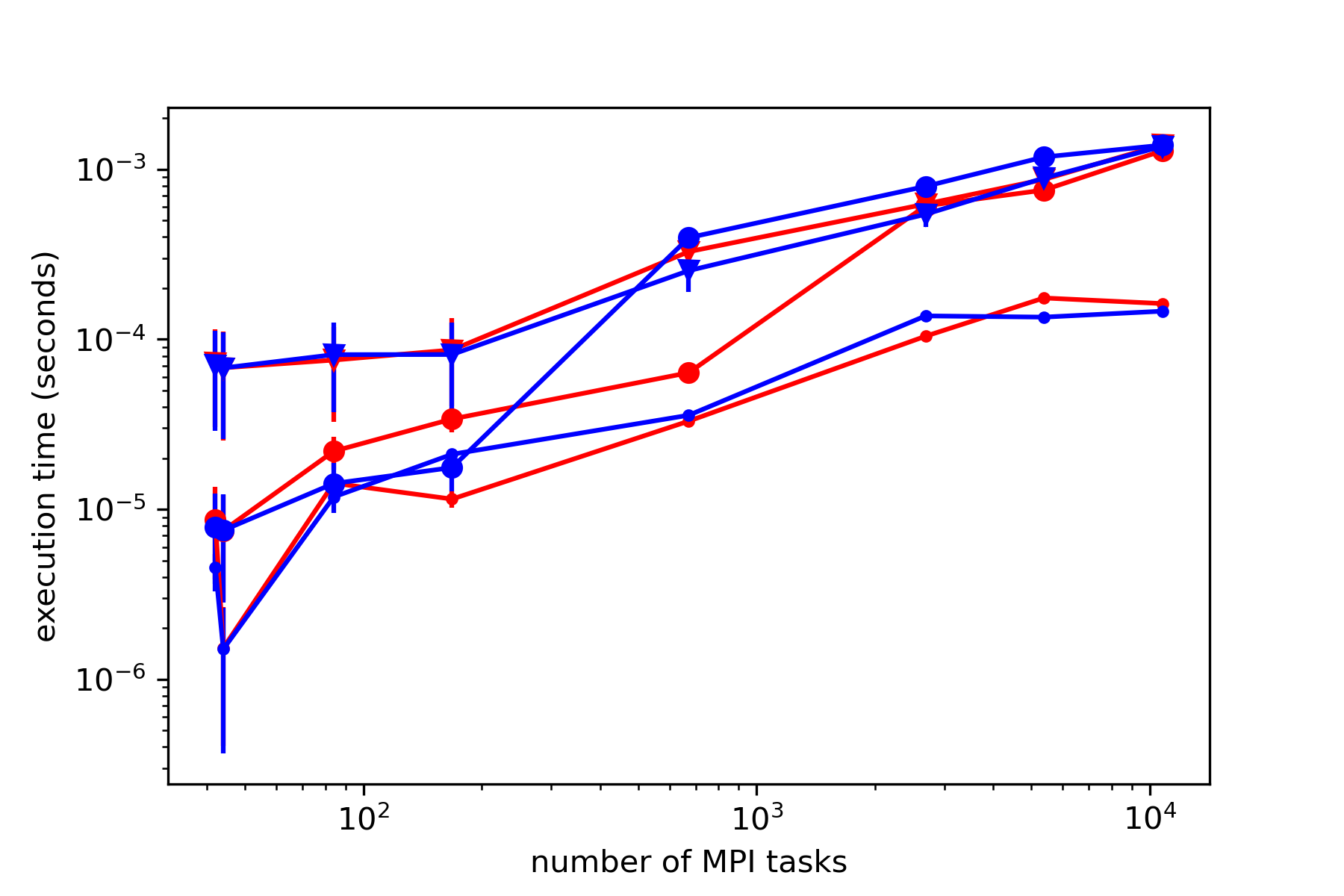}
        \caption{Streaming vector operations}
    \end{subfigure}
    \caption{Comparison of the SUNDIALS MPI-parallel only vector, versus the SUNDIALS MPIPlusX vector with the serial node-local vector used underneath. Each line is the mean execution time of all reduction operations (left) or all streaming operations (right) respectively for a specific local vector length. For most cases the performance difference is negligible, i.e., within performance threshold. In one reduction case and one streaming case, the MPIPlusX vector is perceptibly slower, although the difference is only 3 and 1.3 times (respectively) the magnitude of the performance threshold.}
    \label{fig:mpiplusx-vs-parallel}
\end{figure*}

To demonstrate the low overhead of the MPIPlusX vector, we also conducted tests comparing the standalone SUNDIALS MPI-parallel vector implementation, and the MPIPlusX vector with the serial node-local vector. Again, these tests performed each individual vector operation on random data and the timings were averaged.
The tests were run with 42 MPI tasks per Summit node and 1, 2, 4, 16, 64, 128, and 256 nodes with local vector lengths of $10^3$, $10^4$, and $10^5$ per task\editmade{\footnote{The MPI version used was IBM spectrum-mpi 10.3.1.2-20200121.}}. We define the performance threshold as the minimum execution time of all operations of the same type (streaming or reduction) for the MPIPlusX vector at each length and number of tasks.
The results displayed in Figure \ref{fig:mpiplusx-vs-parallel} show that the overhead associated with the MPIPlusX vector is negligible in most cases. However, with a local vector length of $10^4$, we observed one reduction case and one streaming case where the MPIPlusX vector time was larger than the performance threshold. In these cases, the performance difference was 3 and 1.3 times the magnitude of the error threshold respectively.

\section{Matrix and Solver Structures}
\label{s:matrix-solver}


As described above, SUNDIALS employs an \editmade{object-oriented} approach to algebraic solvers, assuming only that matrix and linear and nonlinear solver implementations supply a minimum set of functionality in order to be used underneath the SUNDIALS time integration modules.
\editstrike{In the GPU-specific context, therefore, so long as the simulation data resides in GPU memory, matrix and solver implementations that similarly store their data and perform calculations on the device may be used for GPU-based SUNDIALS simulations. Thus, since }
\editmade{The} existing set of matrix-free Krylov iterative linear solvers provided by SUNDIALS (GMRES, FGMRES, BiCGStab, TFQMR, and PCG) rely only on vector implementations to both store and operate on simulation data, these solvers may immediately leverage the GPU-based vector implementations discussed in Section \ref{s:vectors}, as well as user-defined vectors.
Additionally, both of the existing nonlinear solvers provided by SUNDIALS, Newton and accelerated fixed-point, rely on vector and linear solver implementations to store and operate on simulation data, and thus they may leverage corresponding GPU-based implementations of those classes.

To complement these iterative linear solvers, we have constructed a new \emph{direct} linear solver object, and corresponding matrix object, for GPU-based calculations.  These leverage NVI\-DIA's batched sparse QR factorization method provided by the cuSOLVER library \cite{cusolver2019manual} and are accordingly named SUNLinear\-Solver\_cuSolverSp\_batchQR and SUNMatrix\_cuSparse, respectively.  As the former primarily serves as a thin wrapper for the SUNDIALS integrators' calls to cuSolverSP routines, we focus our discussion on the latter SUNMatrix\_cuSparse module.  This matrix interface supports the cuSPARSE compressed-sparse-row (CSR) matrix format, as well as a unique low-storage format for block-diagonal matrices of the form
\[
  \mathbf{A}=\begin{bmatrix}
    A_1 & 0 & \cdots & 0\\
    0 & A_2 & \cdots & 0\\
    \vdots & \vdots & \ddots & \vdots\\
    0 & 0 & \cdots & A_n\\
  \end{bmatrix}.
\]
Here, all of the block matrices, $A_j$, are square and share the same sparsity pattern, as encountered in the \editmade{submodel} SUNDIALS use case described in Section \ref{s:considerations} (evolving many small, independent systems within a larger application).  This shared sparsity pattern is leveraged through storing only a single copy of the CSR integer indexing arrays for a generic block $A_i$, resulting in a significant memory savings when using a large number of blocks.  Furthermore, since the sparsity pattern within the QR factorization may also be shared between blocks, this shared sparsity pattern is leveraged within the cuSOLVER library.

For matrix-vector product operations that occur within SUNDIALS, this SUNMatrix\_cuSparse module includes two versions of the sparse matrix-vector product operation.  For matrices in the cuSPARSE CSR matrix format, we directly call NVIDIA's \texttt{cusparseSpMV} routine to perform the multiply.  However, for matrices stored in our custom low-storage format, we provide our own implementation of the low-storage matrix-vector product that leverages the block-diagonal structure and shared sparsity pattern.

\editmade{As with the SUNDIALS-provided GPU vectors, the cuSPARSE matrix and linear solver implementations require that the data is coherent and accessible on the GPU as a one-dimen\-sional array. A set of utility routines are provided to allocate both CSR and block-CSR matrices in device memory and to access the underlying data arrays for the nonzero entries in these matrices.}  \editmade{These} matrix and linear solver implementations may optionally be used with CUDA streams, so long as the same stream is provided to the vector, matrix, and linear solvers that will be used together.
Additionally, more nuanced control over the kernel parameters \editmade{is} used when launching the CUDA kernels, as discussed in Section \ref{s:vectors}.

\section{Using SUNDIALS on CPU+GPU Systems}
\label{s:usingsundials}




\editmadebl{In this section, we discuss how to properly use SUNDIALS on CPU+GPU systems. Interested readers and prospective users should refer to Chapter 6 of the CVODE documentation \cite{cvodeDocumentation} for specific details.}

When using SUNDIALS with GPUs, the integrator control logic is executed on the CPU, and, whenever SUNDIALS is in control of the program, simulation data resides in whatever memory space the vector or matrix object dictates. As such, the general structure of a user's calling program in the GPU-accelerated use case is largely the same as in the CPU-only case\editmadebl{; in fact, all of the calls to integrator (e.g., CVODE) functions are the same}. Rather than using class implementations and providing user-defined functions targeting CPU hardware, the user should utilize GPU-enabled objects and define any user-supplied function to perform calculations on the device as much as possible. \editmadebl{When compiling SUNDIALS with specific GPU features enabled, the produced libraries will have both the CPU and GPU symbols available. Therefore, it is possible to mix CPU objects with GPU objects, but this generally should be avoided for performance reasons expanded upon below.} To summarize, for SUNDIALS simulations on CPU+GPU systems, a user should \editmade{generally}:
\begin{enumerate}
  \item \editmadebl{Compile SUNDIALS with GPU features enabled.}
  \item Utilize a GPU-enabled N\_Vector implementation. \editstrike{Initial data can be loaded on the host, but must be in the device memory space prior to handing control to SUNDIALS.}
  \item Utilize a GPU-enabled SUNNonlinearSolver implementation (if required by the time integrator).
  \item Utilize a GPU-enabled SUNLinearSolver implementation (if required by the nonlinear solver or integrator).
  \item Utilize a GPU-enabled SUNMatrix implementation (if using a matrix-based linear solver).
  \item \editmade{Supply user-defined callback functions that compute the necessary quantities using GPU acceleration.} Examples of these functions include the ODE right-hand side function(s) or the DAE residual function, the Jacobian evaluation function, or the preconditioner setup and solve functions. \editmade{These callbacks must be host functions, and can internally launch GPU kernels.}
\end{enumerate}
\editmade{Two key considerations must be made when using SUNDIALS with GPUs: (1) the user must ensure data coherency between the CPU-host and GPU-device, and (2) for optimal performance it is critical to minimize data movement between the host and device. For this latter point, it is recommended to only access data in the device memory space (unless an atypical data partitioning is used) as much as possible. Ideally, data would reside in device memory for the entire duration of the simulation.}

\editmade{SUNDIALS integrators do not internally migrate data from one memory space to another, and the location of the data depends entirely on the N\_Vector and, if applicable, SUNMatrix implementations that are utilized. With the \textit{SUNDIALS-provided} GPU-enabled vectors and matrices, the data is kept resident in the GPU-device memory space.
Thus, when control is passed from the user's calling program to SUNDIALS, simulation data in vector or matrix objects must be up-to-date in the device memory space if using separate host and device memory (when using UVM, this requirement is handled by the underlying memory system)}.
Similarly, when control is returned to the user, it should be assumed that any simulation data in vector and matrix objects are only up-to-date in the device memory space.
To put it succinctly, it is the responsibility of the user's calling program to manage data coherency between the host and device memory spaces unless unified virtual memory is being utilized.
\editmade{Furthermore, in order to achieve good performance, this means that users need to ensure that the callback functions they supply to the integrators only operate on simulation data in the device memory space otherwise excessive memory transfers will be required. In the case of hybrid data partitioning via the SUNDIALS ManyVector, or a custom N\_Vector and SUNMatrix implementation, the considerations are the same, but the details can be more complex. Generally, users just need to be aware that the vector and matrix objects determine where the data must reside and be updated.}

\begin{figure*}[htb!]
    \centering
    \includegraphics[width=0.75\textwidth]{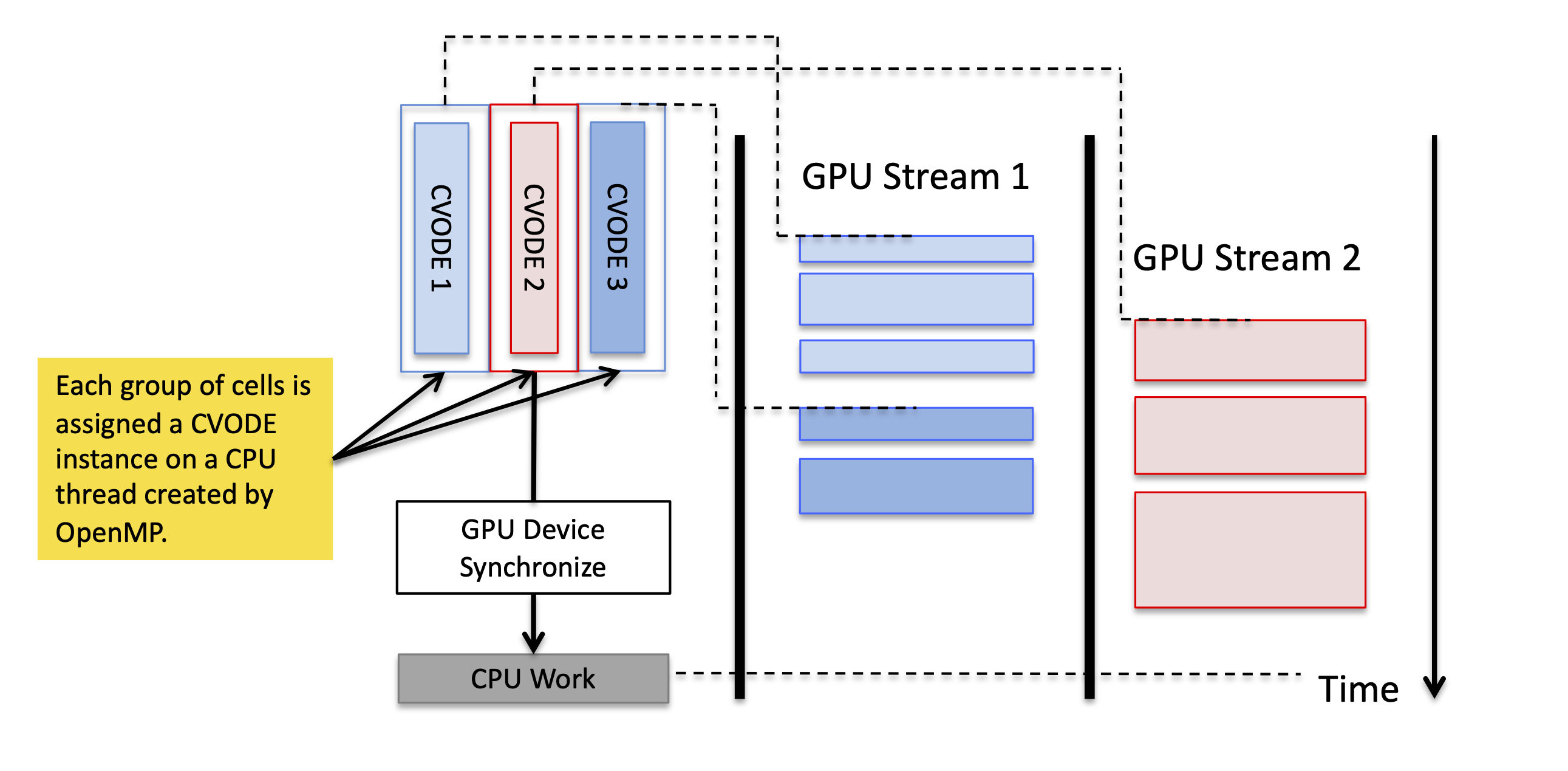}
    \caption{An illustration of using SUNDIALS and hybrid parallelism in combustion applications to integrate many small ODE systems arising in each grid cell of the domain.  In this example, three distinct groups of cells are integrated with CVODE. The groups, each defining a larger ODE system with a Jacobian like Figure \ref{fig:blockdiag_jacobian}, are distributed across CPU threads with OpenMP. On each thread, a distinct and independent CVODE instance solves the larger ODE system. CVODE launches GPU kernels in streams, allowing some threads to operate simultaneously.  Solution data for each group is stored in a distinct, node-local, GPU-enabled N\_Vector. \editmade{\textit{Figure derived from \cite{amrexFigure} and used with permission of the original authors.}}}
    \label{fig:cvodestreams}
\end{figure*}

In the \editmade{submodel} use case, additional parallelism can be obtained by grouping the small independent systems into multiple separate bundles that are integrated concurrently in different GPU streams by unique integrator instances. \editmade{Since integrator function calls are blocking on the host execution thread, each integrator instance must be associated with a different host CPU thread and a corresponding GPU stream to enable concurrent execution.} In this case, the user must also, in addition to the above considerations, ensure that (a) objects are assigned to the appropriate GPU stream so that synchronizations across streams do not occur and (b) the kernel execution policies are set appropriately such that different instances of the integrator may utilize the GPU resources simultaneously. Such a setup with CVODE is illustrated in Figure \ref{fig:cvodestreams} where different CVODE instances are created by and associated with an OpenMP thread on the host. In a loop over multiple grid cells, groups of cells are assigned to a different OpenMP thread and the corresponding CVODE instance which utilizes GPU-enabled objects employing a particular GPU stream. The different CVODE instances are able to integrate their group of cells in parallel to the other CVODE instances and, once that work is complete, return to the pool of cells to integrate another group. This process continues until all the cells have been integrated forward in time.

\section{Demonstration Problem}
\label{s:demonstration}

In this section, we present a demonstration problem that highlights many of the SUNDIALS features for GPUs.
\editmade{The example code is located in the file \wraptt{examples/arkode/CXX_parallel/ark_brusselator1D_task_local_nls.cpp} distributed with SUNDIALS and can be downloaded from the SUNDIALS web page \cite{sundials-web}
or from GitHub \cite{sundials-github}.
}
The problem, a variation of the standard ``brusselator'' test problem from chemical kinetics, solves a one-dimensional advection-reaction system of partial differential equations given by
\begin{align*}
    \frac{\partial u}{\partial t} &= -c \frac{\partial u}{\partial x} + A - (w+1) u + v u^2 \\
    \frac{\partial v}{\partial t} &= -c \frac{\partial v}{\partial x} + w u - v u^2 \\
    \frac{\partial w}{\partial t} &= -c \frac{\partial w}{\partial x} + \frac{B-w}{\epsilon} - w u
\end{align*}
where $u$, $v$, and $w$ are the chemical concentrations, $t$ is time, $x$ is the spatial variable, $c = 0.01$ is the advection speed, $A = 1$ and $B = 3.5$ are the concentrations of chemical species that remain constant over space and time, and $\epsilon = 5 \times 10^{-6}$ is a parameter that varies the stiffness of the system.
We solve the problem on the domain $x \in [0, b]$ with periodic boundary conditions over the time interval $t \in [0, t_f]$.
The initial condition is
\begin{align*}
    u(0,x) &= A + p(x) \\
    v(0,x) &= \frac{B}{A} + p(x) \\
    w(0,x) &= 3.0 + p(x) \\
    p(x) &= \alpha \exp\left(-\frac{(x-\mu)^2}{2\sigma^2}\right),
\end{align*}
where $\alpha = 0.1$,  $\mu = b/2$, and $\sigma = b/4$.
A first order upwind finite difference scheme is used to discretize the spatial derivatives on a uniform mesh with $n_x$ points that are distributed across $n_{px}$ MPI tasks with $n_{xl}$ mesh points per task. As such, there are $3n_{xl}$ ODEs per MPI task.
The discretized system is evolved in time using an implicit--explicit (IMEX) method from the ARKODE package in SUNDIALS in a way that falls under the \editmade{full model} use case described in Section \ref{s:considerations}.
The advection terms are integrated explicitly and the stiff reaction terms are integrated implicitly.
The state variables $u$, $v$, and $w$ are organized into an MPIPlusX vector, and different GPU-enabled node-local vectors are used for evaluating performance.

\begin{figure*}[!htb]
    \centering
    \includegraphics[width=0.8\textwidth]{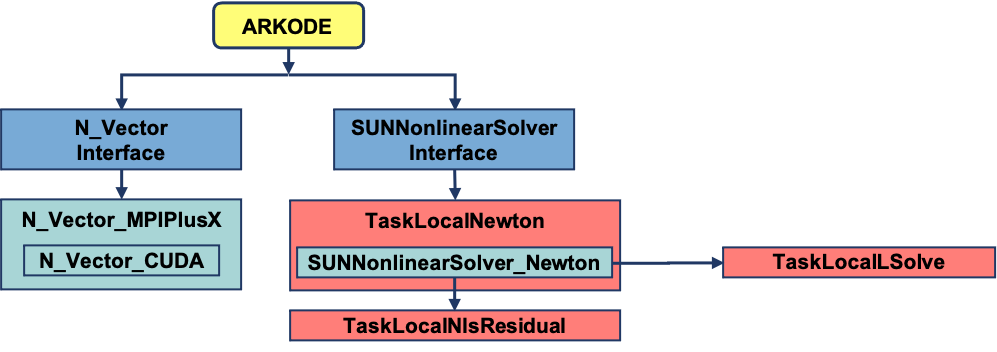}
    \caption{An illustration of the objects utilized in the 1D advection-reaction demonstration code. In this case the example utilizes the CUDA N\_Vector implementation as the node-local vector which is wrapped by the MPIPlusX N\_Vector (left/light green box).  The user code provides a problem-specific implementation of the nonlinear solver class (center/red box) leveraging several the SUNDIALS Newton solver to solve the task-local nonlinear systems.  A custom linear solver method is embedded in the TaskLocalLSolve routine.}
    \label{fig:advrecatorg}
\end{figure*}

Since the reactions are purely local to each mesh point, a custom SUNNonlinearSolver implementation that exploits this locality, and that requires no parallel communication, has been implemented specifically for this example.
On each MPI task the custom solver employs a Newton iteration to solve the $n_{xl}$ implicit systems simultaneously. Each Newton iteration gives rise to a $3n_{xl} \times 3n_{xl}$ block-diagonal linear system that is solved by applying the inverse of each $3 \times 3$ block matrix to the corresponding block vector.
Since each block shares the same structure the operations needed to invert each block matrix are identical; thus, these operations are generated offline with a symbolic Gauss-Jordan method and then are embedded in the demonstration code \cite{wilcoxGaussJordan}. Figure \ref{fig:advrecatorg} illustrates the different code objects used in this configuration of the example.
As noted in \cite{gardner2020enabling}, in older versions of SUNDIALS users could not supply a custom nonlinear solver. As such, it would not have been possible to implement the task-local method described above.
Therefore, the example code also has an option to solve all of the $n_x$ implicit systems together by employing the native SUNDIALS Newton SUNNonlinear and GMRES SUNLinearSolver with the problem-specific block linear solver method previously described serving as a preconditioner.

\editmade{In the following subsection we therefore compare results using both of the nonlinear solver approaches above; these require different types of MPI communication and global synchronization. In both solvers, global reduction operations are required for assessing temporal error; these operations occur by CPU-to-CPU communication.  Additionally, point-to-point communication is required for evaluating the explicit advection operator; when using a GPU vector this communication occurs directly between GPUs, otherwise the transfers occur between CPUs. For the task-local Newton solver, the only additional communication consists of CPU-to-CPU global reduction operations to assess the success or failure of the ensemble of task-local nonlinear solves; if any one solve fails then the entire time step is recomputed using a smaller step size.  For the global Newton solver, global CPU-to-CPU reduction operations are additionally required for each Newton iteration and for all GMRES iterations therein however, no additional point-to-point communications are required due to the task locality of the stiff reaction terms.}

For all configurations of the problem, RAJA is utilized to write the problem-specific code, i.e.~the code specific to the demonstration problem similar to what a user would write when employing SUNDIALS, not code internal to SUNDIALS. Some examples of problem-specific code are the custom nonlinear solver, the preconditioner, the advection operator, and the reaction operator.
Using RAJA for the problem-specific code allows us to switch between different forms of on-node parallelism with minimal code duplication in the demonstration problem; this approach is found in many large-scale applications targeting exascale machines.
It is important to note that using RAJA for the problem-specific code does not necessitate using the RAJA node-local vector, as we will demonstrate in the following sections.

\subsection{Performance on Summit}

To provide a robust look at the GPU performance of SUNDIALS, we now present a weak-scaling study using the 1D advection-reaction problem described above. Nine different configurations of the problem were run.
We define a configuration as a combination of a nonlinear solver option (task-local or global) and a node-local vector: serial, CUDA, CUDA with unified virtual memory (CUDA-UVM), RAJA targeting CUDA, and OpenMP with target offloading (OpenMPDEV).
For brevity we will refer to these configurations as solver+vector,~ e.g. global+serial refers to the global nonlinear solver and the serial node-local vector. All of the runs for this study were conducted on the Summit supercomputer with the goal of maximizing the usage of the relevant computational resources on each node. \editmade{The same versions of XL, Spectrum MPI, CUDA, and RAJA used for the vector tests in Section \ref{ss:vectors-perf} were utilized for this weak-scaling study.}
For configurations which utilize the NVIDIA V100 GPUs, the problem is distributed so that there is one MPI task per GPU, resulting in 6 MPI tasks per node. \editmade{This MPI parititioing results in primarily GPU-GPU communication over the NVIDIA NVLink interconnect, which has a peak bandwidth of 50 GB/s \cite{summitOLCF}.}
The MPI-only configurations use 40 MPI tasks per node, thus using 40 of the 42 available CPU cores per node while maintaining the same $n_{xl}$ for all MPI tasks. \editmade{With this MPI partitioning, most communications are core--core and via shared-memory copies; the peak memory bandwidth for these transfers is 170 GB/s \cite{summitOLCF}.}
All configurations use a fixed spatial resolution $b/n_{x} = 1.67 \times 10^{-5}$, and the domain size, $b \in \{10, 40, 160, 640, 2560\}$, is increased with the number of MPI tasks to keep $n_{xl}$ constant.
The configurations are run with the number of nodes $\in \{1, 4, 16, 64, 256\}$, thus $n_x$ ranges from $6 \times 10^{5}$ to $1.536 \times 10^8$, i.e.~$n_{xl} = 10^5$ for the GPU-enabled configurations and $n_{xl} = 1.5 \times 10^4$ for the MPI-only configurations.
The simulations with the task-local solver are run to $t_f = 10.0$ while the global configurations are run to $t_f = 3.0$ since they take significantly longer.
Every configuration is run 25 times, and the timing results are averaged.



\begin{figure}[!htb]
    \centering
    \includegraphics[width=0.4\textwidth]{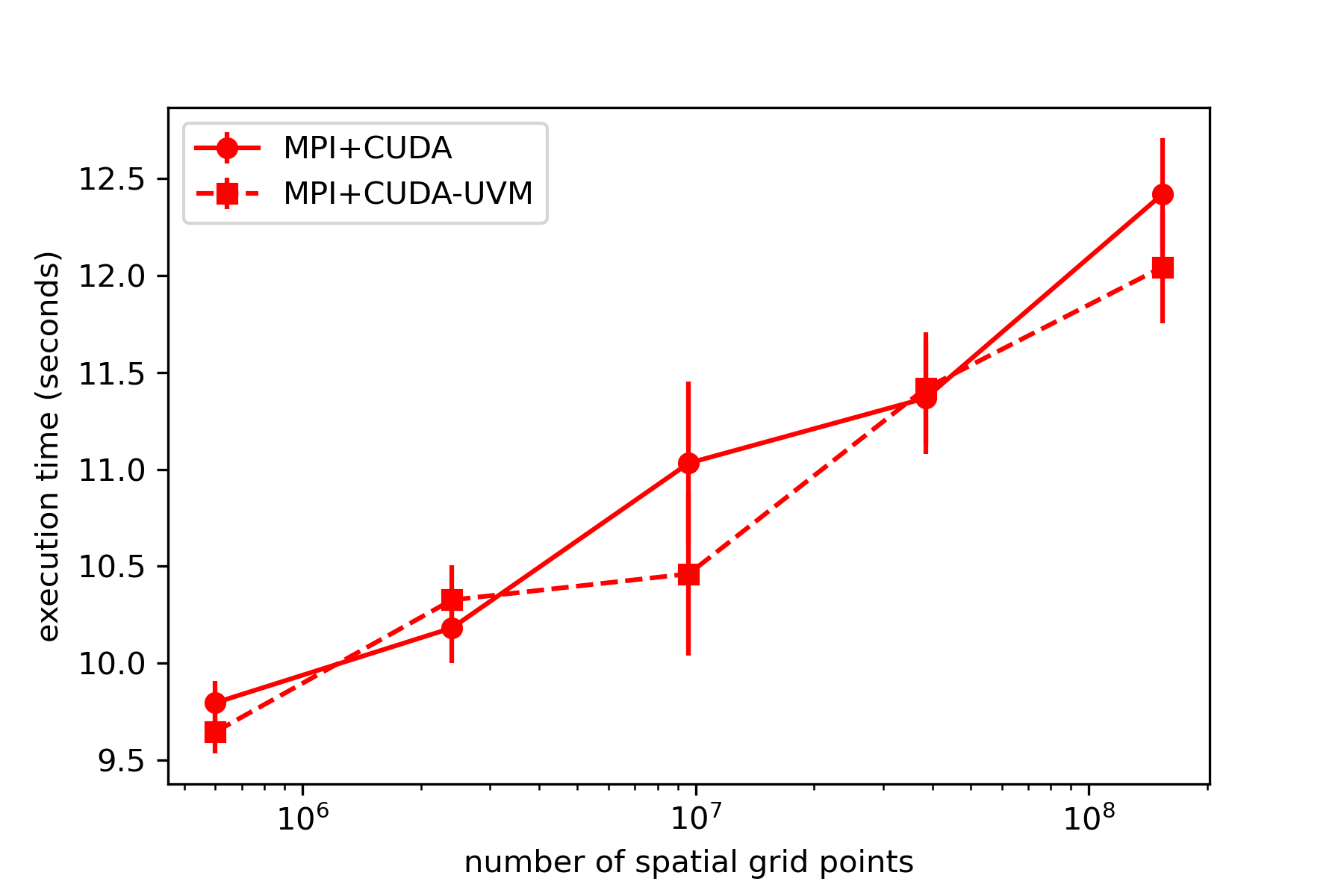}
    \caption{Weak scaling of the 1D advection-reaction problem using the task-local nonlinear solver with the CUDA node-local vector with distinct host and device memory versus with unified virtual memory. The error bars represent the standard deviation of the MPI+CUDA execution times. Since we carefully manage data movement in the demonstration problem, the performance difference is always negligible (within one standard deviation).}
    \label{fig:demo-more}
\end{figure}

\begin{figure*}[!htb]
    \centering

    \begin{subfigure}[b]{0.49\textwidth}
        \includegraphics[width=\textwidth]{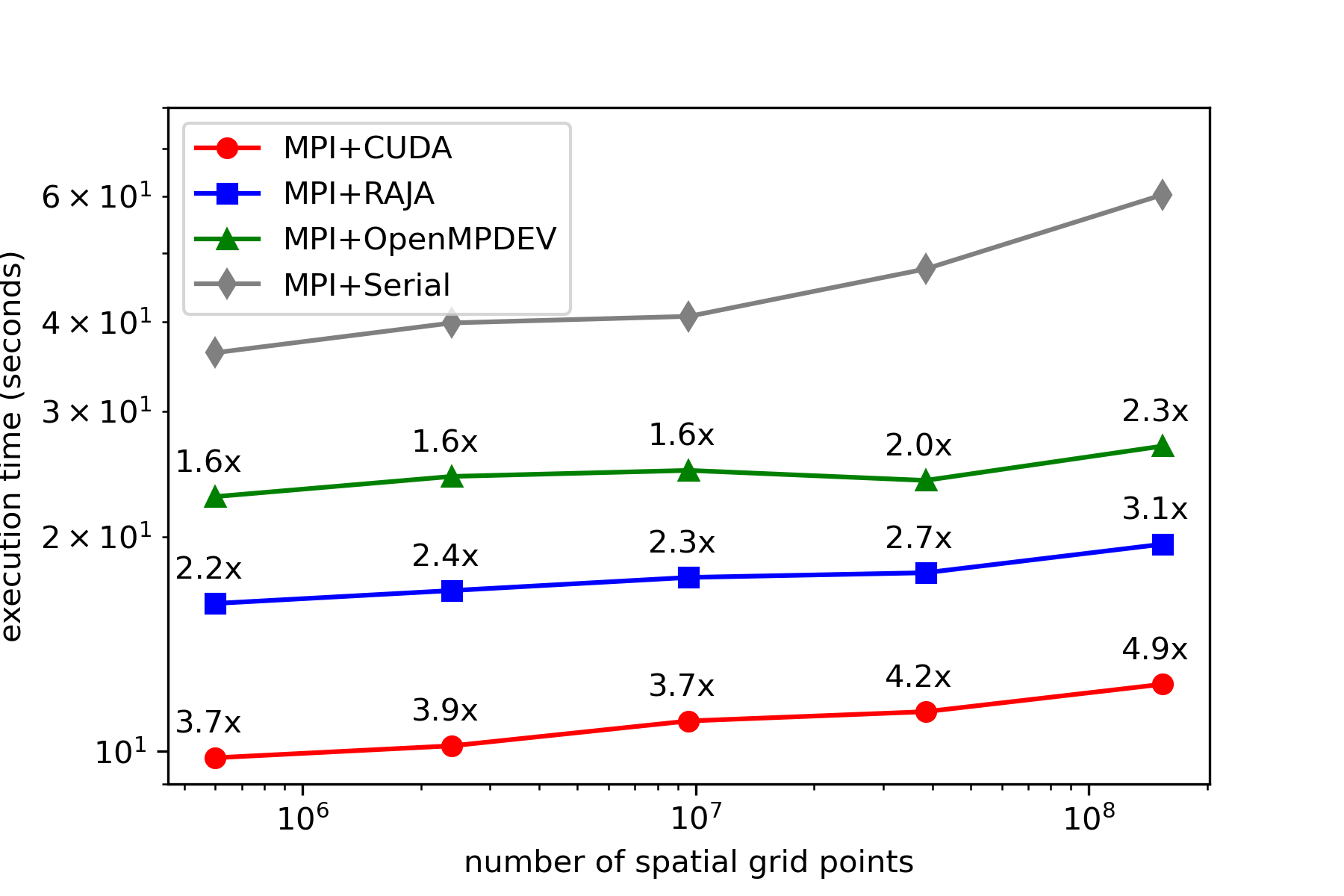}
        \caption{Task-local Newton}
    \end{subfigure} \hfill
    \begin{subfigure}[b]{0.49\textwidth}
        \includegraphics[width=\textwidth]{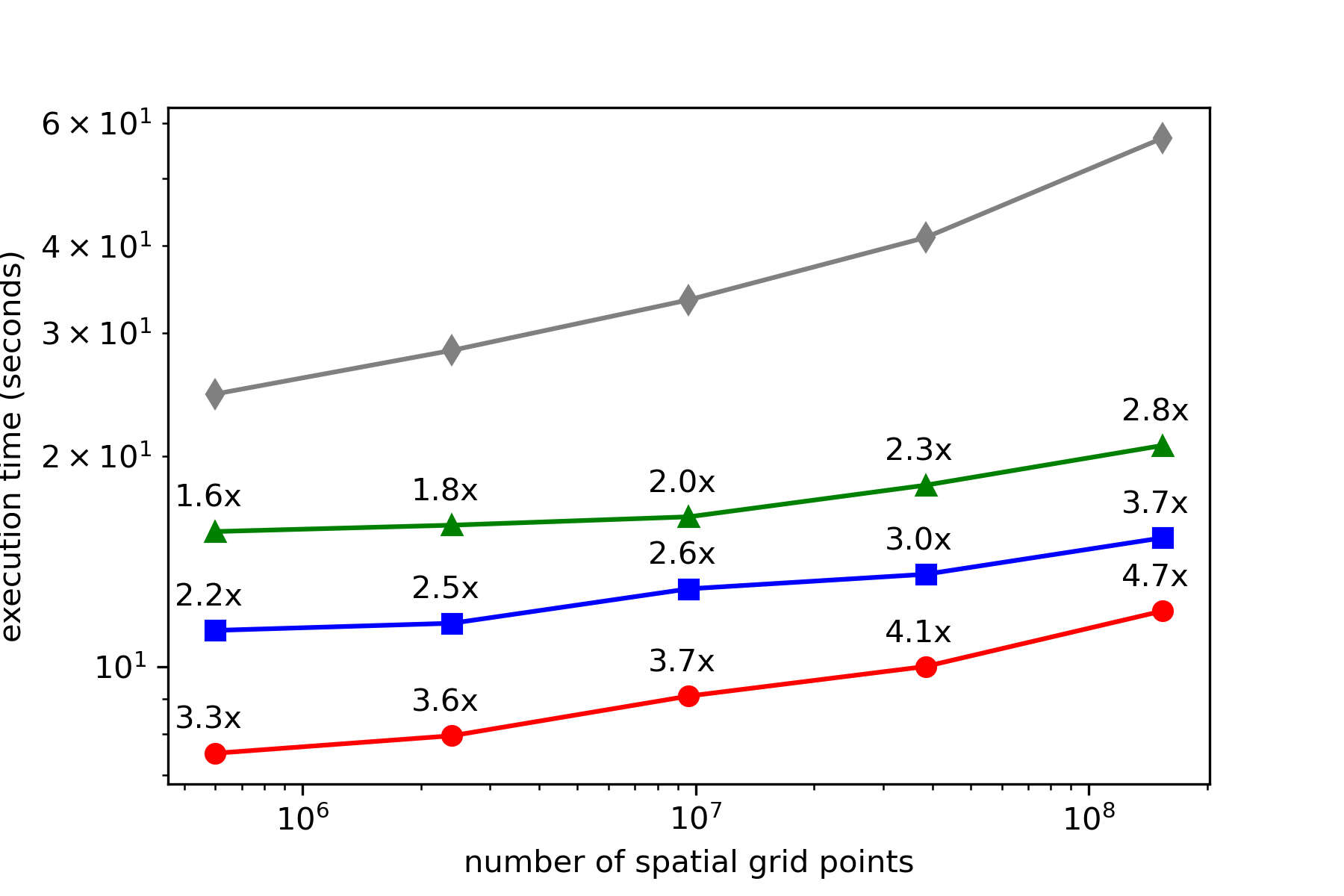}
        \caption{Global Newton}
    \end{subfigure}

    \caption{Weak scaling of the 1D advection-reaction problem using the MPI+X vector with different node-local vectors and two different algebraic solvers: the task-local Newton nonlinear solver that employs direct linear solves (left), and the global Newton nonlinear solver that employs preconditioned GMRES (right). Task-local configurations are run until $t_f = 10$, while the global configuration was run until $t_f = 3$ since the simulation takes significantly longer. The largest problem uses 256 nodes and all 6 GPUs per Summit node. Data points are annotated with the speedup over the configuration with the MPI+Serial vector.}

    \label{fig:demo-weakscale}
\end{figure*}

Figures \ref{fig:demo-more} and \ref{fig:demo-weakscale} plot the weak-scaling results for all of the configurations.
When using the same nonlinear solver, the runs using the CUDA node-local vector are the fastest, besting the runs using the RAJA vector and OpenMPDEV vector. \editmade{The performance difference between the CUDA and RAJA vectors is more significant than in the vector tests discussed in Section \ref{ss:vectors-perf}; this is largely due to the fact that the demonstration problem is reliant on a small subset of the vector operations, e.g. N\_VLinearSum and N\_VScale, where the performance difference is larger than the mean over all operations presented above.}

\editmade{Comparing} the task-local+CUDA to the task-local+serial runs, the results show a speedup of 3.7x -- 4.9x.
That is, to solve the problem with the same number of spatial grid points and on the same domain, the task-local+CUDA configuration is 3.7x -- 4.9x faster than the task-local+serial.
Using this same definition of speedup, the global+CUDA configuration is 3.3x -- 4.7x faster than the global+serial configuration.
At the largest problem size, the task-local+serial configuration achieves only 60\% parallel efficiency while the task-local+CUDA, task-local+RAJA, and task-local+OpenMPDEV achieve 79\%, 83\%, and 85\% respectively. The global configurations generally scale less efficiently; at the largest problem size the global+serial achieves 43\% parallel efficiency and the global+CUDA, global+RAJA, global+OpenMPDEV achieve 63\%, 74\%, and 75\% efficiency respectively.
In addition, the results also demonstrate that the performance difference between runs that use UVM and those that do not (see Figure \ref{fig:demo-more}) are within the standard deviation of the measured execution times with the CUDA vector and no UVM. As such, the performance difference can be considered to be negligible.

\begin{figure*}[!htb]
    \centering

    \begin{subfigure}[b]{0.49\textwidth}
        \includegraphics[width=\textwidth]{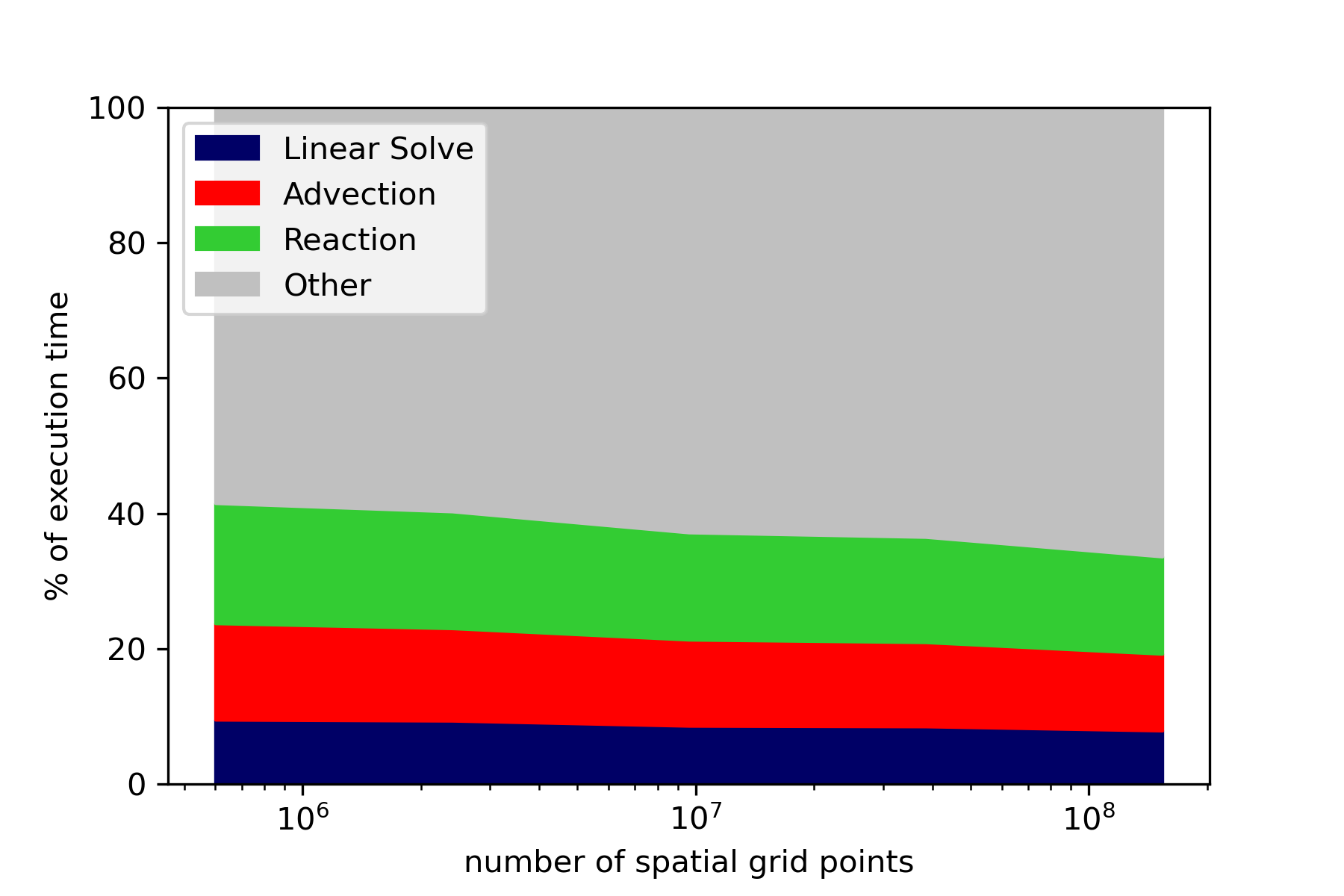}
        \caption{Task-local CUDA}
    \end{subfigure} \hfill
    \begin{subfigure}[b]{0.49\textwidth}
        \includegraphics[width=\textwidth]{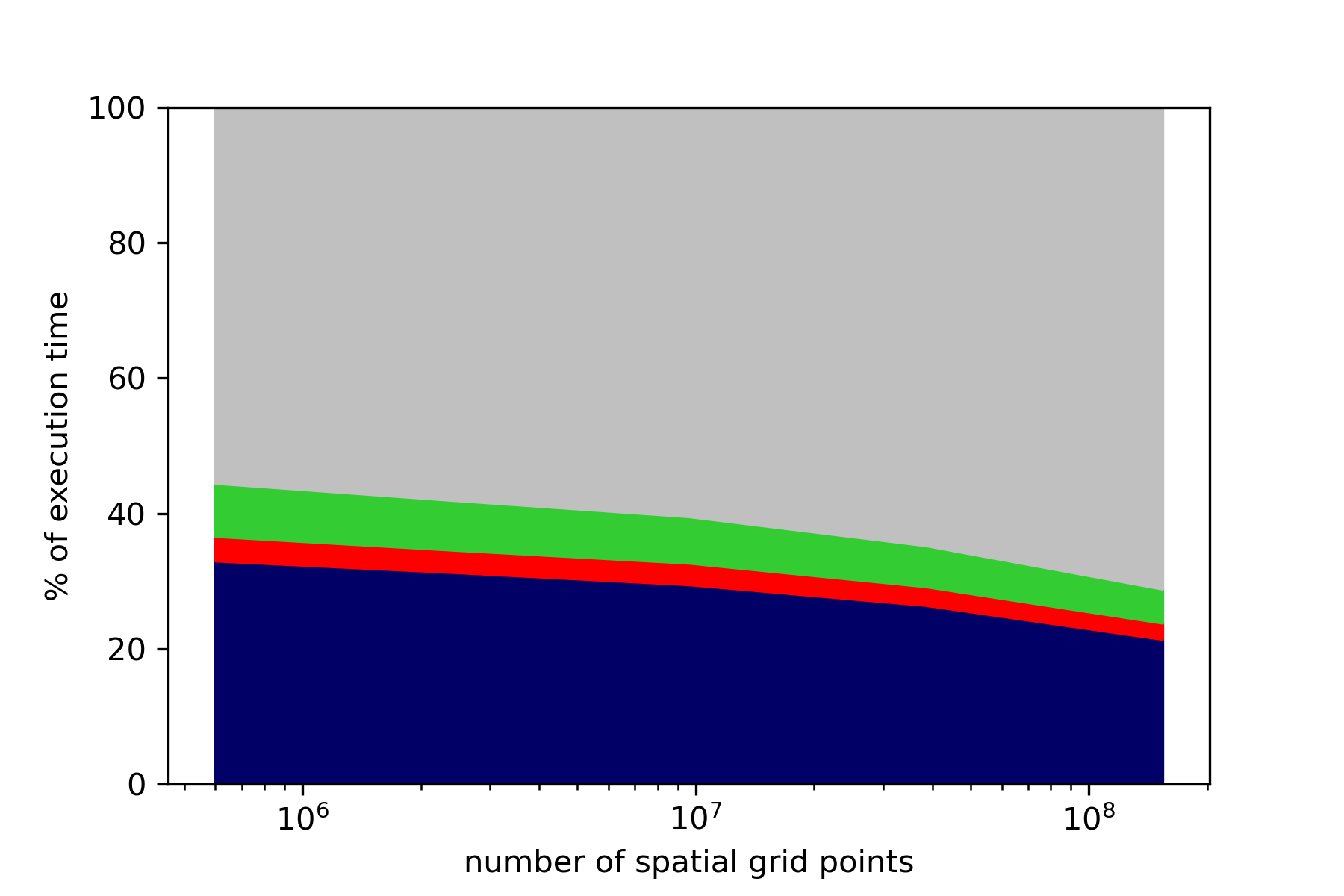}
        \caption{Task-local Serial}
    \end{subfigure}

    \caption{\editmade{Breakdowns of the execution time for the task-local+CUDA (left) and task-local+serial (right) 1D advection-reaction problem configurations demonstrate that choice of solver can have significant impacts on where time is spent within a simulation. In the task-local+CUDA case, the average total execution times were 10.18s, 11.03s, and 11.36s for the $10^6$, $10^7$, and $10^8$ problem sizes respectively. For the task-local+serial case, the average times were 39.90s, 40.76s, and 47.49s.}}

    \label{fig:demo-breakdown}
\end{figure*}

To provide a better understanding of where the execution time is spent and where speedup is obtained we compare (see Figure \ref{fig:demo-breakdown}) the timing of four different code regions for the task-local+CUDA configuration and task-local+serial configuration.
The explicit advection operator category includes the time spent communicating values at process boundaries and the finite difference stencil computations.
The implicit reaction operator includes the computation time only, since there is no communication required. The linear solve category includes computing the Jacobian, setting up the linear system, and then solving it.
Finally, the other category includes everything else. Specifically ``other'' includes core integrator functionality such as computing and controlling errors, and the nonlinear solve excluding the linear solve phase --- all of which involve vector operations (streaming and reduction).

\editmade{From these} results we make several observations.
First, the task-local+CUDA spends a similar amount of time in the explicit advection operator as the task-local+serial\editmade{, but as a percentage of the total execution time, it is greater in the CUDA case}. Based on our analysis of the kernel launch latency in Section \ref{ss:vectors-perf}, we can attribute this behavior \editmade{in part} to the overhead of launching the CUDA kernels for the finite difference stencil computation. \editmade{An additional factor might be that the task-local+CUDA involves primarily intranode GPU--GPU communication while task-local+serial primarily involves faster core--core communication}.
The results also show that the linear solve and other regions of the task-local+CUDA configuration are significantly faster than in the serial case. Thus, we can state that the overall speedup is resulting from both faster vector operations as well as a faster linear solve.

Summarizing, our weak-scaling study on Summit shows the importance of SUNDIALS algorithmic flexibility as it allows us to employ the much faster and more scalable task-local nonlinear solver. Moreover, these experiments demonstrate the low overhead of the additional infrastructure SUNDIALS required for flexibility and GPU-enabled computing as well as the significant performance benefit of using the GPU-enabled features of SUNDIALS.

\subsection{Performance on Frontier EAS}

In addition to the above weak-scaling study conducted on Summit, we also execute tests on the Tulip early access system (EAS) for the upcoming Frontier exascale supercomputer.
We again utilize the 1D advection-reaction problem, but only two configurations are tested: the task-local nonlinear solver with the CUDA vector and the HIP vector. Simulations are run to $t_f = 10.0$, with $n_x \in \{10^5, 10^6, 10^7\}$, $b \in \{10, 100, 1000\}$, and only one MPI task is employed.
The CUDA vector configuration targets a NVIDIA V100 GPU while the HIP vector targets either an AMD MI60 or MI100 GPU.

\begin{figure}
    \centering

    \includegraphics[width=0.4\textwidth]{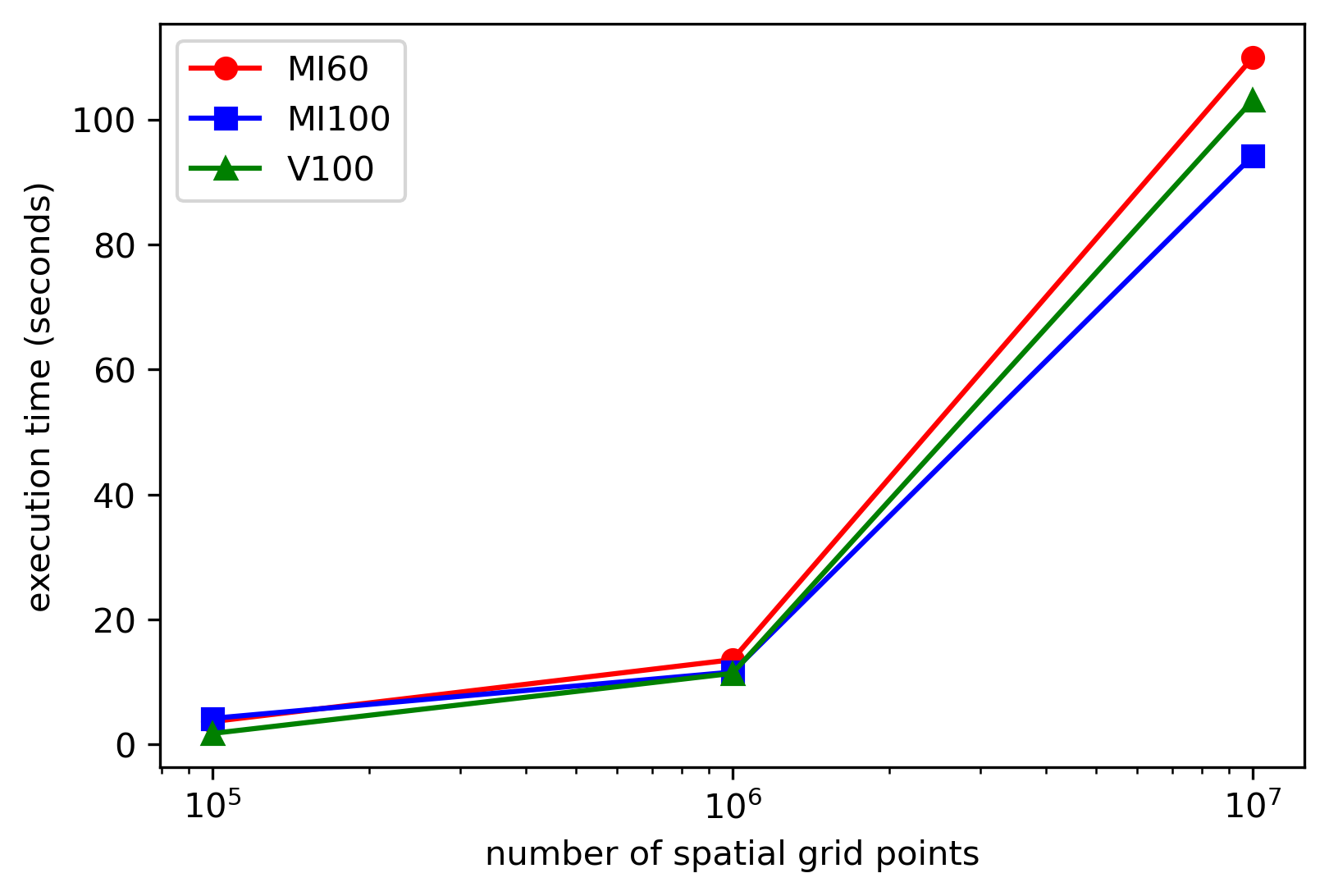}

    \caption{Execution time of the 1D advection-reaction problem using a single MPI task and a single GPU (NVIDIA V100, AMD MI60, or AMD MI100) on the Tulip EAS for Frontier. The V100 runs are the fastest for the smallest problem size ($n_x = 10^5$), results are similar across all GPUs when $n_x = 10^6$, and for the largest problem size ($n_x = 10^7$) the MI100 runs are 10\% faster than the V100 runs and 17\% faster than the MI60 runs.}
    \label{fig:demo-tulip}
\end{figure}

Figure \ref{fig:demo-tulip} shows the results from these tests. The task-local+CUDA configuration is the fastest for the smallest problem size ($n_x = 10^5$), but the task-local+HIP targeting the AMD MI100 is 10\% faster than the task-local+CUDA for the largest problem size ($n_x = 10^7$).
As demonstrated in the Summit weak-scaling study, the bulk of the execution time is spent inside of the ``other'' category, i.e.~mostly vector operations. The SUNDIALS vector operations accounting for most of the runtime for this particular problem are memory-bound; thus, the aforementioned performance can be explained by examining both the theoretical memory bandwidth of the different GPU devices and the achieved memory bandwidth.
Profiling of the demonstration problem shows that the N\_VLinearSum operation is the most costly integrator operation in terms of cumulative execution time, thus in Table \ref{tab:demo-bandwidth} we list the mean achieved memory bandwidth of N\_VLinearSum (obtained by taking the averaged achieved bandwidth of each call to the operation). The data in this table contextualizes the behavior of the execution time -- for the smallest problem size (when the V100 run is fastest) the achieved memory bandwidth is much higher with the V100 GPU and the CUDA vector, but for the largest problem size (when the MI100 run is fastest) the achieved bandwidth with the MI100 GPU and the HIP vector is greatest, by approximately 12\%, and beyond the  peak theoretical bandwidth of the V100.

\begin{table}[!htb]
    \centering
    \caption{The theoretical peak memory bandwidth of each GPU targeted on Tulip, and the mean achieved memory bandwidth of the most expensive (in execution time) integrator operation (N\_VLinearSum) when running the 1D advection-reaction problem with a single MPI-task/GPU on Tulip. Higher bandwidth is achieved with the CUDA vector and V100 for the smallest problem but with the MI100 and HIP vector for the largest problem.}
    \begin{tabular}{lcccc}
        \toprule
        GPU & theoretical peak (GB/s) & \multicolumn{3}{c}{achieved (GB/s)} \\
        \midrule
        $n_x$ &      & $10^5$ & $10^6$ & $10^7$ \\
        \midrule
        V100  &  900 & 699 & 766 & 814 \\
        MI60  & 1000 & 271 & 674 & 787 \\
        MI100 & 1200 & 212 & 739 & 921 \\
        \bottomrule
    \end{tabular}
    \label{tab:demo-bandwidth}
\end{table}

\section{Conclusions}
\label{s:conclusions}
SUNDIALS follows an \editmade{object-oriented} design that allows extreme flexibility with respect to the underlying data structures and algebraic solvers; in this work we leverage this design to enable efficient GPU-based simulations within the SUNDIALS time integrators.  Specifically, we target two main GPU use cases, (1) \editmade{the full model case where} the time integrator evolves the full application model in time, and (2) \editmade{the submodel case where} a SUNDIALS time integrator evolves many, small, independent systems within a larger application.

\editmade{To} facilitate these use cases, SUNDIALS needed to be equipped with vector data structures to support several GPU programming models. In addition, the SUNDIALS implicit time integrators further need  GPU-enabled algebraic solvers, both iterative and direct.
To these ends, we developed new vector structures based on CUDA, HIP, RAJA (with a CUDA backend), and OpenMP with device offloading, as well as an MPI+X hybrid structure that utilizes MPI for distributed memory parallelism along with node-level acceleration for GPU programming models.
With these GPU-enabled vectors in place, the SUNDIALS iterative algebraic solvers, such as Krylov methods for linear systems and Newton methods for nonlinear systems, as needed in the \editmade{full model} use case, were immediately able to run with GPUs.
Motivated by the \editmade{submodel use case}, we further enabled CUDA-based direct linear solvers, through construction of a SUNDIALS matrix structure that supports both CSR and block-CSR formats, as well as accelerator-level matrix access and arithmetic operations.
This matrix infrastructure will easily support solvers based on other GPU programming models, including HIP and DPC++, as linear solvers based on those models become more available.

\editmade{In} addition, we added a set of utilities to allow easy tuning of GPU kernels and use of application-based memory management infrastructures underneath the SUNDIALS-provided data structures and solvers.
These utilities provide a minimal common interface that the native SUNDIALS structures can leverage to maximize code reuse within SUNDIALS and flexibility for users.

\editmade{Results} given above indicate that the overhead of using the MPI+X vector structure is generally negligible.
An examination of our node-local vectors (minus HIP) both in isolated tests and in a weak-scaling study of an advection-reaction problem on Summit, showed that the CUDA vector performed the best out of the SUNDIALS CUDA-enabled node-local vectors.
Results from the advection-reaction problem also showed 3.3x-4.9x speedup using MPI+CUDA over MPI+Serial depending on the nonlinear solver used.
In tests on the Tulip early access system for Frontier, the HIP vector targeting the AMD MI100 GPUs provided a 10\% performance benefit for the largest problem size over the CUDA vector targeting NVIDIA V100 GPUs largely due to the greater memory bandwidth.

\editmade{These} results show that the additions we have made to the SUNDIALS suite to support GPUs, including the new MPI+X vector infrastructure and flexibility for new algebraic solver classes, have little performance overhead and allow the suite to exploit GPUs for significant performance benefit.
Future work will expand on these capabilities through inclusion of DPC++ based vectors to target Intel GPUs as well as HIP and DPC++ compatible direct solvers.

\editmadebl{Finally, we part with some broader observations based on our personal experiences developing GPU support for SUNDIALS. First, our work suggests that targeting GPUs directly through the vendor specific programming models and languages typically results in better performance at this time. While better performance is excellent, we have also demonstrated that current performance portable GPU programming models (RAJA and OpenMP offloading) can still provide significant speedup compared to using the CPU exclusively. Performance portable GPU programming models also offer sustainability benefits, including a lower maintenance burden and reduced reliance on developers knowing vendor specific language details.
Furthermore, based on our demonstration problem, it seems that a combination of a performance portability layer and vendor specific code could offer the best of both approaches to GPU programming. We advise that other software developers enabling GPU computing in mathematical libraries consider the effort required to implement and support multiple programming models in the context of their projects and weigh the trade-offs of portability against achieving the best possible performance.}

\section*{Acknowledgements}

The authors would like to thank Shelby Lockhart of the University of Illinois for her help in implementing the SUNDIALS vector using OpenMP with device offloading, and Daniel McGreer of the University of California at Santa Cruz for his help in implementing a HIP-based SUNDIALS vector.

This work was performed under the auspices of the U.S. Department of Energy by Lawrence Livermore National Laboratory under Contract DE-AC52-07NA27344. Lawrence Livermore National Security, LLC. LLNL-JRNL-816232. This document was prepared as an account of work sponsored by an agency of the United States government. Neither the United States government nor Lawrence Livermore National Security, LLC, nor any of their employees makes any warranty, expressed or implied, or assumes any legal liability or responsibility for the accuracy, completeness, or usefulness of any information, apparatus, product, or process disclosed, or represents that its use would not infringe privately owned rights. Reference herein to any specific commercial product, process, or service by trade name, trademark, manufacturer, or otherwise does not necessarily constitute or imply its endorsement, recommendation, or favoring by the United States government or Lawrence Livermore National Security, LLC. The views and opinions of authors expressed herein do not necessarily state or reflect those of the United States government or Lawrence Livermore National Security, LLC, and shall not be used for advertising or product endorsement purposes.

This research used resources of the Oak Ridge Leadership Computing Facility, which is a DOE Office of Science User Facility supported under Contract DE-AC05-00OR22725.

\editmade{This work utilized the early-access system Tulip hosted by HPE and supported by the HPE and AMD staff of the Frontier Center of Excellence.}

Funding: This research was supported by the Exascale Computing Project (17-SC-20-SC), a collaborative effort of the U.S. Department of Energy Office of Science and the National Nuclear Security Administration.

\section*{References}
\bibliography{sources}

\end{document}